\UseRawInputEncoding

\documentclass[trackchanges, twocolumn, twocolappendix]{aastex7}

\usepackage{comment}
\usepackage{enumitem}

\newcommand{\rvir}{R$_{\rm vir}$}
\newcommand {\msun}{M$_{\odot}$}
\newcommand {\Msunperyear}{M$_{\odot}$/yr}

\newcommand{\zzero}{$z=0$}

\usepackage{color}
\usepackage{graphics}
\usepackage{amsmath}
\usepackage{multirow}
\usepackage{booktabs}
\usepackage{lipsum}
\usepackage{braket}
\graphicspath{{./}{./}}

\newcommand{\HI}{\ion{H}{1}$\,$}

\newcommand{\LHI}{$\mathcal{L}_{HI}$}

\newcommand{\HIcm}{\ion{H}{1}~21-cm}

\newcommand{\LessActive}{\textit{Less populated}}
\newcommand{\MoreActive}{\textit{More populated}}

\def\rev#1{#1}

\defcitealias{peeples19}{FOGGIE~I}
\defcitealias{corlies20}{FOGGIE~II}
\defcitealias{zheng20}{FOGGIE~III}
\defcitealias{simons20}{FOGGIE~IV}
\defcitealias{lochhaas21}{FOGGIE~V}
\defcitealias{lochhaas23}{FOGGIE~VI}
\defcitealias{wright24}{FOGGIE~VII}
\defcitealias{Acharyya24}{FOGGIE~VIII}

\defcitealias{trapp25b}{FOGGIE~XIII}

\usepackage{ragged2e}

\begin{document}

\title{FOGGIE: Figuring Out Gas \& Galaxies In Enzo XII. The Formation and Evolution of Extended \HI\ Galactic Disks and Warps with a Dynamic Circumgalactic medium}
\shorttitle{Evolution of HI Disks in FOGGIE}

\author[orcid=0000-0001-7813-0268,sname='North America']{Cameron W. Trapp}
\affiliation{Center for Astrophysical Sciences, William H.\ Miller III Department of Physics \& Astronomy, Johns Hopkins University, 3400 N.\ Charles Street, Baltimore, MD 21218}
\email[show]{ctrapp2@jhu.edu}  

\author[0000-0003-1455-8788]{Molly S.\ Peeples}
\affiliation{Space Telescope Science Institute, 3700 San Martin Dr., Baltimore, MD 21218}
\affiliation{Center for Astrophysical Sciences, William H.\ Miller III Department of Physics \& Astronomy, Johns Hopkins University, 3400 N.\ Charles Street, Baltimore, MD 21218}
\email{molly@stsci.edu}

\author[0000-0002-7982-412X]{Jason Tumlinson}
\affiliation{Space Telescope Science Institute, 3700 San Martin Dr., Baltimore, MD 21218}
\affiliation{Center for Astrophysical Sciences, William H.\ Miller III Department of Physics \& Astronomy, Johns Hopkins University, 3400 N.\ Charles Street, Baltimore, MD 21218}
\email{tumlinson@stsci.edu}

\author[0000-0002-2786-0348]{Brian W. O'Shea}
\affiliation{Department of Computational Mathematics, Science, and Engineering, Michigan State University, East Lansing, MI, US}
\affiliation{Department of Physics and Astronomy, Michigan State University, East Lansing, MI, US}
\affiliation{Facility for Rare Isotope Beams, Michigan State University, East Lansing, MI 48824, USA}
\affiliation{Institute for Cyber-Enabled Research, 567 Wilson Road, Michigan State University, East Lansing, MI 48824}
\email{bwoshea@msu.edu}

\author[0000-0003-1785-8022]{Cassandra Lochhaas}
\affiliation{Space Telescope Science Institute, 3700 San Martin Dr., Baltimore, MD 21218}
\affiliation{Center for Astrophysics, Harvard \& Smithsonian, 60 Garden St., Cambridge, MA 02138}
\affiliation{NASA Hubble Fellow}
\email{clochhaas@cfa.harvard.edu}

\author[0000-0002-1685-5818]{Anna C.\ Wright}
\affiliation{Center for Computational Astrophysics, Flatiron Institute, 162 Fifth Avenue, New York, NY 10010}
\email{awright@flatironinstitute.org}

\author[0000-0002-6804-630X]{Britton D. Smith}
\affiliation{Institute for Astronomy, University of Edinburgh, Royal Observatory, EH9 3HJ, UK}
\email{Britton.Smith@ed.ac.uk}

\author[orcid=0009-0000-7559-7962,sname='Saeedzadeh']{Vida Saeedzadeh}
\affiliation{Center for Astrophysical Sciences, William H.\ Miller III Department of Physics \& Astronomy, Johns Hopkins University, 3400 N.\ Charles Street, Baltimore, MD 21218}
\email{vsaeedz1@jh.edu}  

\author[0000-0003-4804-7142]{Ayan Acharyya}
\affiliation{INAF - Astronomical Observatory of Padova, vicolo dell’Osservatorio 5, IT-35122 Padova, Italy}
\email{ayan.acharyya@inaf.it}

\author[0000-0001-7472-3824]{Ramona Augustin}
\affiliation{Leibniz-Institut f{\"u}r Astrophysik Potsdam (AIP), An der Sternwarte 16, 14482 Potsdam, Germany}
\email{raugustin@aip.de}

\author[0000-0002-6386-7299]{Raymond C.\ Simons}
\affiliation{Department of Engineering and Physics, Providence College, 1 Cunningham Sq, Providence, RI 02918 USA}
\email{rcsimons@providence.edu}

\begin{abstract}
 
Atomic hydrogen (\HI) is an important component of gas in and around galaxies and forms extended disk-like structures well beyond the extent of starlight. Here we investigate the properties and evolution of extended \HI\ disks that emerge in six Milky Way-mass galaxies using cosmological zoom-in simulations from the Figuring Out Gas \& Galaxies in Enzo (FOGGIE) suite. We focus on the formation, evolution, and morphology of extended gaseous disks that emerge in all six systems. We find that median \HI\ column densities drop sharply at the disk edge, with mean column densities outside the disk dominated by dense ($N_{\rm HI}\sim10^{19} \rm{cm}^{-2}$), clumpy structures. All systems have significant misaligned features (warps or polar rings) at some point in their evolution; however, their frequencies, lifetimes, and origins vary significantly. We find that the morphologies of the FOGGIE disks are correlated with properties of their Circumgalactic Medium (CGM).
\rev{We place these systems along a continuum based on how populated their CGMs are with \HI\ relative to their central disk. All systems kinematically settle similarly by $z=0$. The \LessActive\ systems tend to form coherently rotating, thin, extended disks while the \MoreActive\ systems do not.}
\rev{Location on this continuum is independent of disk and halo mass, implying a relation to local environmental factors.}
Our results indicate a connection between CGM content and disk formation that is not yet fully understood. A second paper investigates observational aspects of these structures.

\end{abstract}


\section{Introduction} \label{sec:introduction}

The extended, outer disks of galaxies act as an important interface between the Circumgalactic Medium (CGM) and the inner regions of the disk, where the bulk of star formation occurs. Recent advances in interferometric radio arrays have enabled the creation of high-sensitivity, high-resolution \HI\ maps \cite[e.g.,][]{Walter_2008-THINGS}. Along with increasingly sensitive single-dish observations \cite[e.g.,][]{wang2024-FEASTS_I}, these techniques continue to lower observable gas surface density limits and can characterize \HI\ at larger and larger radii. A robust theoretical understanding of how the extended \HI\ disks form, evolve, and relate to both the inner disk and CGM is vital to understanding the latest observations and answering the question of how galaxies get their gas.

Misaligned features in the outer regions of disk galaxies are commonly observed. Warps, in which there is a continuous component of gas at larger radii inclined relative to the inner disk, are relatively common, with many local galaxies, including our own, showing signatures of warps \cite[e.g.,][]{burke57-mw_warps,reshetnikov98-warp_stats}. Various mechanisms have been proposed to explain the formation of these warps, including bending modes in a self-gravitating disk \citep{sparke98-WarpBendingModes}, misaligned dark matter halos \cite[e.g.,][]{dubinski95-DarkHaloWarps}, subhalo interactions \cite[e.g.,][]{kim14-WarpsInteractions, Semczuk20-WarpsInteractions}, \rev{and accretion from a misaligned hot-halo \citep{Sriram2025-IdealizedWarps}}. Polar ring systems, in which a distinct component of gas orbits at a large inclination from the disk plane, are less commonly observed, although many examples still exist \cite[e.g.,][]{whitemore90-PolarRingObs, Finkelman12-PolarRingObs}. Polar rings are typically thought to originate from merger events \citep{bekki97-PolarRingMergers, bournaud-PolarRingFormation}, tidal accretion from interacting subhalos \citep{schweizer83-PolarRingAccretion, reshetnikov97-PolarRingAccretion}, or potentially, cold-mode accretion from filaments \citep{stanonik09-PolarRingFilaments}.

Cosmological hydrodynamical simulations are an important means by which we can attempt to understand the formation and evolution of gaseous disks. Simulations have studied numerous aspects of disks, including disk settling, thin disk formation, and the transition from bursty to stable star formation \cite[e.g.,][]{muratov15,sparre16,furlanetto22, Gurvich_2022-FireDiskSettling, furlanetto22,Hafen22,donkelaar22}. While these three evolutionary milestones are correlated, they do not necessarily happen in the same order if at all \citep{hopkins23-DiskTransitions}. Some simulations also find a correlation between the transition from bursty to stable star formation and the virialization of the inner CGM \citep{Stern21}, in which the material within 0.1 \rvir\ becomes thermalized and net accretion slows down. This implies that accurate modeling of material in the extended disks and inner CGM is important to our understanding of all of these disk formation processes. Warps and polar rings have also been seen in other cosmological simulations of galaxies, although their prevalence and lifetimes vary \cite[e.g.,][]{roskar10, gomez17, Semczuk20-WarpsInteractions, smirnov24, 2026MNRAS.545f1963L}.

This paper is the first in a series of two papers in which we study the extended disks, polar rings, and warps that emerge naturally from the six Milky-Way-like galaxies in the Figuring Out Gas \& Galaxies in Enzo (FOGGIE) suite of simulations \citep{peeples19}. The FOGGIE halos were selected to yield galaxies at $z = 0$ that resemble the Milky Way in total mass ($M_{\rm halo} \simeq 0.5-1 \times 10^{12}~\rm M_{\odot}$).
However, the halos were not tuned to produce extended \HI\ disks or larger disk-like structures. Even so, all of them have extended \HI\ during their evolution. We also observe warped disks and polar rings that emerge naturally, aided by FOGGIE's high-resolution gas refinement schemes.

This paper focuses on the theoretical characterization of these extended \HI\ disks. To that end, we investigate how the extent and morphology of these extended disks evolve, how the angular momentum components of the inner and outer disks relate to each other, the formation and evolution of polar rings and warps, and how these disks settle and form thin disks over cosmological time. We find two distinct categories of disks, those with CGMs that are \LessActive\ with \HI\ and those with CGMs that are \MoreActive\ with \HI. These local environmental properties ($R \lesssim 2 R_{\rm vir}$) have significant impacts on disk morphology, particularly their disk structures at late times.

The second paper in this series \citep{trapp25b} compares directly with recent extended \HI\ observations, creating synthetic \HIcm\ datacubes and investigating the effects of sensitivity, spatial resolution, and missing short baselines in interferometric studies. It will hereafter be referred to as \citetalias{trapp25b}. \rev{Similar to recent comparisons to simulations \citep[e.g.][]{marasco25:mhongoose-sim-comp,lin2025-FeastsCompWithSimulations}, we find the FOGGIE galaxies show more low column density \HI\ than seen in observations on average; however, it varies greatly between systems and can be sensitive to the effect of missing short baselines.}

\rev{For the remainder of this paper, when we refer to disks, we are referring to the gaseous disk component, unless otherwise mentioned.} In Section~\ref{sec:simulations} we briefly describe the FOGGIE simulations. In Section~\ref{sec:hi_disk_definition}, we discuss how we separate the disk from the CGM. In Section~\ref{sec:hi_profiles} we describe the extended \HI\ profiles of the six halos. In Section~\ref{sec:disk_evolution} we discuss how these systems grow, kinematically settle, and form thin disks over cosmological time. In Section~\ref{sec:tempest_blizzard_comparison}, we provide a more in-depth comparison of two systems that both form polar rings and warps, although with different mechanisms and characteristic timescales. We additionally investigate numerical resolution considerations in Appendix~\ref{sec:simulated_resolution}. Appendix~\ref{sec:other_halo_time_evolutions} looks at other halos' time series, and Appendix~\ref{sec:disk_definition_appx} discusses the details of our disk definition.

\begin{deluxetable*}{lcccccccccc}
\tabletypesize{\scriptsize}
\tablecaption{Key parameters for the six FOGGIE Halos considered in this study at redshift $z=0$.}
\label{tab:halo_masses}
\tablehead{
\colhead{Halo} & 
\colhead{$M_{\rm 200}$} & 
\colhead{$M_{*,\mathrm{disk}}$} & 
\colhead{$M_{\rm HI,200}$} & 
\colhead{$M_{\rm HI,disk}$} & 
\colhead{$R_{\rm 200}$} &
\colhead{$R_{\rm HI}$} & 
\colhead{$H_{\rm HI}$} & 
\colhead{$\langle \dot{M_{*}}/M_{*}\rangle$} & 
\colhead{$|\langle dM/dz \rangle|$} \\
\colhead{} &
\colhead{[10$^{12}$ M$_{\odot}$]} & 
\colhead{[10$^{10}$ M$_{\odot}$]} & 
\colhead{[10$^{10}$ M$_{\odot}$]} & 
\colhead{[10$^{10}$ M$_{\odot}$]} & 
\colhead{[kpc]} & 
\colhead{[kpc]} & 
\colhead{[kpc]} & 
\colhead{[Gyr$^{-1}$]} & 
\colhead{[10$^{11}$ M$_{\odot}$]}
}
\startdata
\textbf{\textit{Less Populated}} \\
Tempest   & 0.50 &  4.02 & 1.28 & 1.27 & 168.3 & 21.8 & 1.25 & 0.05 & 1.64 \\
Maelstrom & 1.01 &  7.24 & 2.72 & 2.36 & 211.9 & 24.2 & 0.69 & 0.07 & 4.14 \\
Blizzard  & 1.14 &  9.02 & 2.79 & 2.33 & 220.3 & 22.0 & 0.69 & 0.06 & 3.81 \\
Cyclone   & 1.69 &  1.74 & 3.88 & 3.17 & 246.8 & 38.2 & 2.36 & 0.05 & 6.23 \\
Hurricane & 1.71 & 17.20 & 3.73 & 2.94 & 252.0 & 33.3 & 0.69 & 0.10 & 6.80 \\
Squall    & 0.80 & 10.48 & 0.73 & 0.50 & 195.9 & 16.8 & 1.81 & 0.10 & 2.14 \\
\textbf{\textit{More Populated}} \\
\enddata
\tablecomments{ The masses shown are either the mass within $R_{200}$ (e.g. $\mathbf{M_{\rm HI,200}}$), or the mass within our disk definition (e.g. $\mathbf{M_{\rm HI,disk}}$). $R_{200}$ is the radius enclosing an average density of 200$\times$ the critical density of the universe at $z=0$. $\mathbf{R_{HI}}$ is calculated as the radius at which the mean \HI\ column density falls below $1.25\times10^{20}$ cm$^{-2}$, a commonly used observational contour \citep[e.g.][]{wang16-HISizeMass,bluebird20:chilesVI}. $\mathbf{H_{HI}}$ is the mean \HI\ scale height within this radius. This will overestimate the thickness of the disk in systems with coherent misaligned features. $\mathbf{
\langle \dot{M_{*}}/M_{*}\rangle}$ is the mean specific star formation rate between z=0.5-0. $\mathbf{|\langle dM/dz\rangle|}$ is the absolute value of the mean halo growth rate between z=0.5-0. All halos are growing.}
\end{deluxetable*}

\section{FOGGIE Simulations} \label{sec:simulations}

We analyze six zoom-in galaxy formation simulations of Milky Way-mass halos from the FOGGIE suite. The FOGGIE simulations were run using the open-source adaptive mesh refinement code Enzo \citep{bryan14,brummel-smith19}. These simulations are distinct from most other zoom-ins in that they are designed to resolve the CGM through a novel ``forced refinement'' scheme, enabling high spatial resolution in even diffuse gas \citep{peeples19} as well as a ``cooling refinement'' criterion to better resolve thermally unstable gas.

The FOGGIE simulations have been used to study, interpret, and produce observable predictions for a range of CGM and galaxy applications. These include: absorption \citep{peeples19} and emission-line \citep{corlies20,lochhaas25,saeedzadeh25}, observations of the CGM, absorption-line observations of the CGM of the Milky Way \citep{zheng20}, ram pressure stripping of dwarf satellites \citep{simons20}, the kinematic support and dynamical equilibrium of circumgalactic gas \citep{lochhaas21, lochhaas23}, the origin and evolution of stellar halos in galaxies \citep{wright24}, the spatially-resolved evolution of metallicity in galaxies  \citep{Acharyya24}, the physical and chemical properties of gas clumps in the CGM \citep{Augustin25}, the evolution of angular momentum \citep{simons25}, and the accretion through the CGM towards galaxies (C.\ Lochhaas et al., in prep).

The FOGGIE simulations and their forced refinement resolution scheme were first introduced in \citet{peeples19}. We specifically use the second generation of simulations, first introduced in \citet{simons20}. These simulations have a comoving box size of 100$h^{-1}$ Mpc with a 256$^{3}$ root grid. The FOGGIE simulations use a forced-refined box with a length of 288 kpc comoving on a side. Every gas cell within this fixed box centered on and moving with the main galaxy is forced to a minimum comoving spatial resolution (1100 pc comoving). At each timestep of the simulations, cells within this box are allowed to further refine based on density.  

\rev{FOGGIE is distinct among most other cosmological simulations in that it focuses its computational elements on resolving the CGM, as opposed to denser regions within the central disk.
Standard cosmological zoom-in simulations, such as NIHAO \citep{nihao-i}, FIRE-2 \citep{hopkins18-FIRE2}, FIRE-3 \citep{fire3}, DC Justice League \citep{applebaum21}, AURIGA \citep{2017MNRAS.467..179G}, and Vela \citep{Vela2014MNRAS.442.1545C}, focus resolution on regions of high gas density, such as the central disk, interstellar medium, and satellite galaxies. Since the CGM is relatively low-density, it typically has poor spatial resolution in these schemes. These simulations will typically have better resolution within the central disk, while FOGGIE will typically have better resolution within the CGM.}

Starting with \cite{simons20}, the FOGGIE simulations implemented a ``cooling refinement'' criterion, which refines cells such that their size is smaller than the cooling length (sound speed $\times$ cooling time). The density and cooling refinement criteria are allowed to resolve up to 274 pc comoving. This scheme ensures that a large fraction (well over 90\%) of the CGM mass is resolved according to this cooling refinement criterion. In Sec.~\ref{sec:resolution_criteria}, we investigate how well material in the disk is resolved by these criteria. Briefly, we find that the density and cooling refinement are important; however, the cooling refinement criterion is triggered more often than the density refinement. The median cell mass in the extended disks is comparable to other state-of-the-art cosmological simulations ($\sim10^{4}$ M$_{\odot}$), while the disk-halo interface is much better resolved, even only a few cells away from the disk.

\rev{The FOGGIE simulations include density and metallicity-dependent cooling and a metagalactic background \citep{haardt12} using the {\sc Grackle} chemistry and cooling library \citep{smith17}, including self-shielding of gas owing to \HI\ opacity at $z\leq15$ \citep{emerick19}. The code solves a non-equilibrium six species chemical reaction network at run time, tracking \HI, \ion{H}{2}, \ion{D}{1}, \ion{D}{2}, \ion{He}{1}, \ion{He}{2}, \ion{He}{3}, and e$^{-}$. All metal species are grouped in a single field, allowing for metallicity-dependent cooling assuming ionization equilibrium and solar abundances.}

\rev{Simulations are run with Enzo's native star formation and supernovae thermal feedback schemes \citep{cen_ostriker_06}. This feedback is underpowered, in part leading to these systems having overmassive and overly concentrated stellar profiles. Star formation occurs in dense gas with a converging flow. Gas is turned into star particles in proportion to the local gas mass, with a minimum star-particle mass of 1000 M$_{\odot}$ at high redshift, ramping up to 10,000 M$_{\odot}$ between $z=2$ and $z=1$ (see \citealt{wright24} for a more thorough discussion of stellar feedback and star formation prescriptions). Dark Matter particle mass is $1.39\times10^{6}~\rm{M}_{\odot}$.}

\rev{There are an increasing number of other cosmological simulations focusing on resolving the CGM. GIBLE \citep{ramesh24-gible} is another cosmological zoom-in simulation that focuses resolution on the CGM using the same galaxy formation model as IllustrisTNG. It reaches median spatial resolutions of 75 pc to 700 pc (comoving) within the CGM, compared with FOGGIE's fixed spatial resolution of 274 pc (comoving). Unlike FOGGIE, GIBLE includes black hole/AGN feedback and magnetic fields, and shows a drop off in mass/spatial resolution within the central galaxy. The {\sc MEGATRON} simulations have also added additional CGM refinement \citep{cadiou25}, including non-equilibrium cooling and radiation transport. These simulations include refinement on jeans length and, similar to FOGGIE, cooling length. Due to increased computational costs, they have only been run to $z\sim4$. ENGAWA \citep{lucchini26} is a recent suite of cosmological zoom-in simulations that similarly focuses additional CGM resolution for four halos from the Auriga project \citep{grand17} and TNG50 \citep{pillepich19-TNG50}. Starting at redshift $z=0.3$, these simulations maintain a fixed volume refinement of 200 pc resolution out to a radius of 100 kpc and throughout the disk, allowing for a focus on the CGM and the disk halo interface at late times. FOGGIE has high CGM resolution at from high to low redshifts, allowing for a full characterization of a system's evolution over time.}

In this study, we analyze all six halos in the second generation of FOGGIE simulations. All halos were selected to be Milky Way-like at $z=0$ in terms of their mass and merger history (see \citealt{wright24} for more details). After $z=2$, there are no major mergers (with the exception of Squall, which experiences a 2:1 merger at $z=0.7$). All snapshot outputs were saved with a high temporal cadence (every 5.4 Myr). To identify subhalos, dark matter halos within the zoom region are identified with the {\sc ROCKSTAR} halo finder \citep{Behroozi2013a}, which uses a friends-of-friends algorithm in combination with temporal and 6D phase-space information. Virial quantities $R_{\rm vir}$ and $M_{\rm vir}$ for each halo are also calculated by {\sc ROCKSTAR} using the redshift-dependent $\rho_{\rm{vir}}$ of \citep{bryan98}. Merger histories for each halo are assembled with {\sc CONSISTENT-TREES} \citep{Behroozi_2013b} and halo properties are collated across time using {\sc TANGOS} \citep{pontzen2018}.

\begin{figure*}[ht!]
\plotone{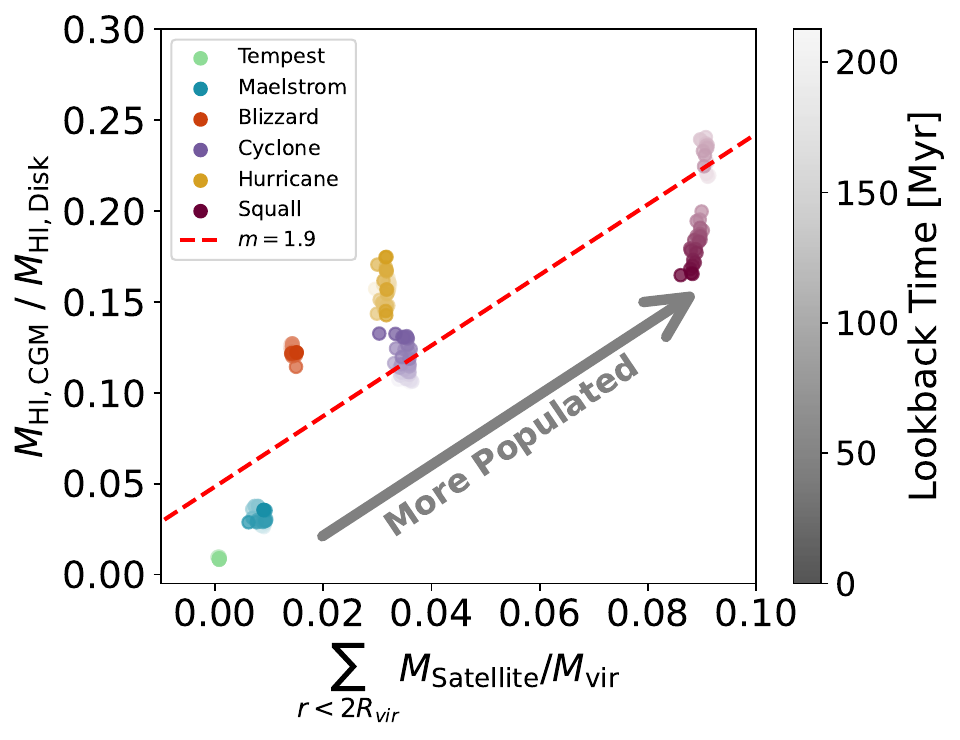}
\caption{
Satellite mass within $2 R_{\rm vir}$ relative to the main halo virial mass, versus the \HI\ mass in the CGM relative to the \HI\ mass of the disk over the last $\sim$200 Myr. There is a clear correlation between the relative amount of \HI\ in the CGM with the nearby satellite mass at that time. \rev{The \HI\ mass within 1 \rvir\ from satellites, as identified by {\sc ROCKSTAR} was not included in the y-axis.} Systems on the bottom left of this plot are in \LessActive\ local environments, while systems in the top right are in \MoreActive\ local environments. We subdivide our sample into two categories based on their position along this relation. Tempest, Maelstrom, and Blizzard comprise the \LessActive\ systems, while Cyclone, Squall, and Hurricane comprise the \MoreActive\ systems. We emphasize that this is a continuum. Blizzard is an edge case, as the most \HI\ rich of the \LessActive\ systems. Cyclone moving down the continuum at late times as well, which may reflect other galactic properties.
}
\label{fig:classification_quantification}
\end{figure*}

\begin{figure*}[ht!]
\begin{centering}
\includegraphics[width= .825\textwidth]{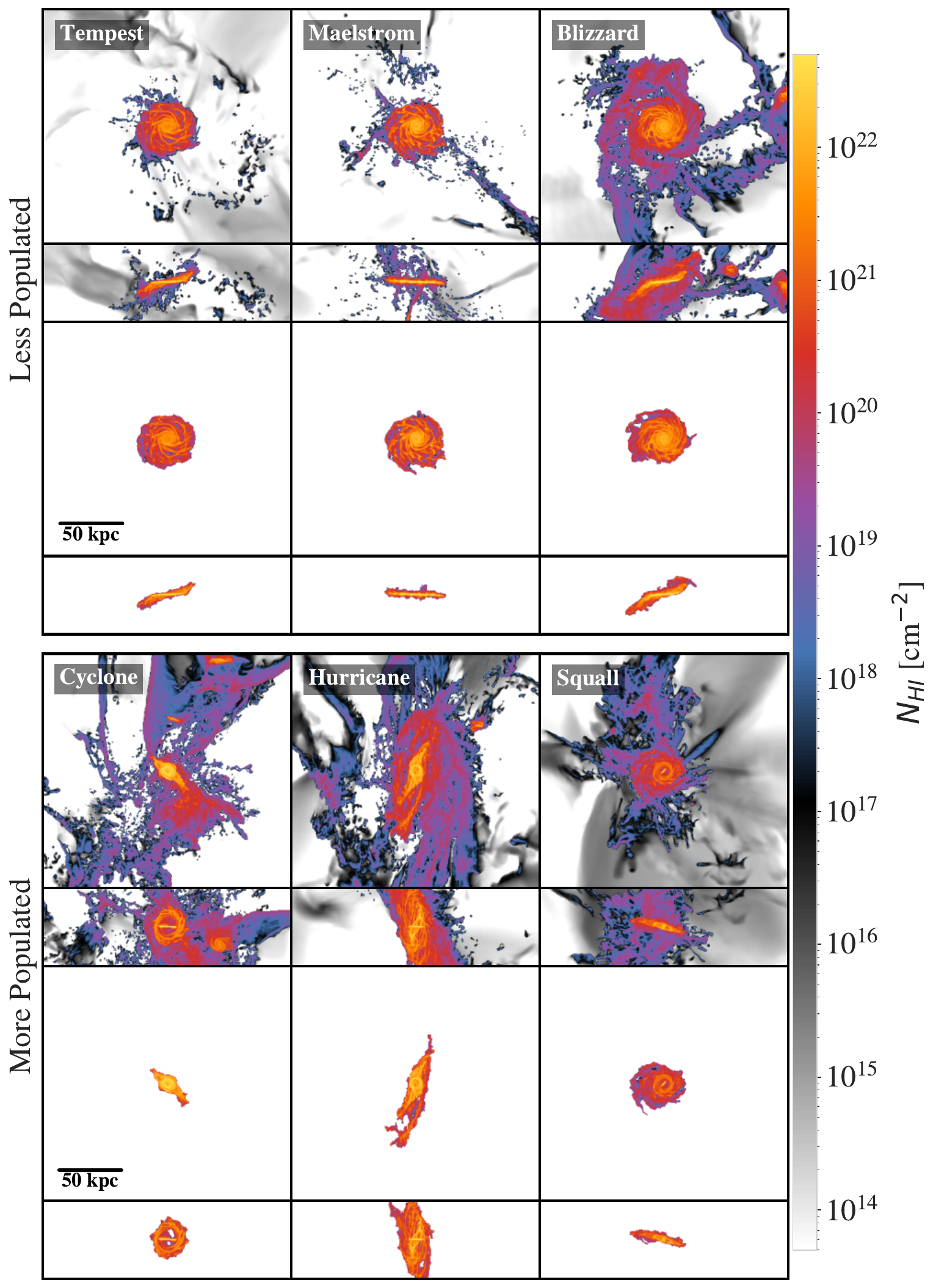}
\caption{
Face- and edge-on \HI\ column densities within $R_{\rm vir}$ for the six halos considered in this study at redshift \zzero. We subdivide our halos into two distinct categories (\LessActive~and \MoreActive) based on the amount of \HI\ in the CGM. The top row of each category shows the \HI\ column densities for all gas, while the bottom row shows the projections only including gas in our disk definition (see Sec.~\ref{sec:hi_disk_definition}). As can be seen, this categorization correlates strongly with the structure of the disk.  The color map transitions to grey scale at $N_{\rm HI}<10^{17.5}$, which is the high end of sensitivity for the latest \HI\ interferometry surveys \citep{deBlok24-Mhongoose}.
\label{fig:hi_projections}}
\end{centering}
\end{figure*}

\section{Defining the \HI\ Disk} \label{sec:hi_disk_definition}

Separating the \rev{gaseous} disk of a galaxy from the CGM is important for understanding how the two evolve and interact. Simple geometric cuts based on the characteristic radius and height of the gaseous disk (a cylinder) are often used to make this distinction. The selection of this height cutoff is important, as too thick a cutoff will include significant extra-planar gas, while too thin a cutoff will miss features such as disk flaring and misaligned gas. Similar concerns arise when attempting to define the CGM with geometric solids. These issues become increasingly important when there are significant misaligned components, such as the warps and polar rings frequently seen in the FOGGIE simulations and observed galaxies \citep[e.g.,][]{reshetnikov98-warp_stats, Finkelman12-PolarRingObs}.

To address the limits of simple geometric cuts, we present a novel way to identify the disk based on a 3D clump finding technique. We classify a ``disk'' or central galaxy as the most massive, contiguous distribution of \HI\ gas above a certain density cut. We set an initial \HI\ number density cut ($n_{\rm cgm}=6.5\times10^{-3} ~\rm{cm^{-3}}$) to separate CGM and disk gas such that a 1 kpc thick column of gas would have a column density of more than $2 \times 10^{19}$ cm$^{-2}$ at $z=0$. To account for increasing densities at higher redshifts and for differences in galactic environments, we modify this cutoff as follows:
\begin{equation} \label{eq:disk_cutoff}
    n_{\rm cut} = \frac{n_{\rm cgm}   + \gamma\sigma_{\rm cgm}}{a^{3}}~,
\end{equation}
where $\sigma_{\rm cgm}$ is the volume-weighted standard deviation of the \HI\ number density of the CGM gas as identified by the initial density cut ($n_{\rm cgm}$) and within the virial radius ($R_{\rm vir}$), $a$ is the scale factor of the Universe at the given time, and $\gamma$ is a unitless constant. The choice of $\gamma$ allows for control over how much the environment affects this cut. For systems with less material in their CGMs, particularly Tempest and Maelstrom, the choice of $\gamma$ does not make a significant difference at low redshifts. At higher redshifts and for systems with more material in their CGMs (Cyclone, Hurricane, Squall), a large value is needed to consistently identify the central disk. We select $\gamma=100$ as it gives consistent results across all redshifts and systems.

The choices of $\gamma$ and $n_{cgm}$ are somewhat arbitrary, with values chosen experimentally to robustly isolate the central galaxy across the redshift regimes covered in this study. Given the sharp drop off in column density at the disk edges seen in the FOGGIE simulations from $10^{21}~\rm{cm}^{-2}$ to $10^{18}~\rm{cm}^{-2}$ at $z=0$ (see Sec.~\ref{sec:desc_of_extended_disks}), this disk definition does not vary strongly with the specific choice of $n_{cut}$, as shown in Appendix~\ref{sec:disk_definition_appx}.   
The factor of $a^{-3}$ is necessary, as otherwise the disk definition includes too much material in the CGM at high redshifts for all systems. 

Since we do not want our \rev{gaseous} disk definition to exclude low-density regions in the disk associated with stellar feedback or dynamical effects, we fill holes within the disk in two steps. The first fills any true topological holes (fully enclosed), such as a feedback bubble smaller than the disk height. To fill ``donut'' holes that completely pierce the disk (e.g., spiral structure or chimney effects from feedback), we apply a binary closing operation \citep{petros87-BinaryClosing} with a kernel size of 7 kpc. In brief, this operation performs a binary dilation on the disk mask, followed by a binary erosion. Around the edges of the disk mask, these two operations will cancel each other out. If the dilation step fills any holes, they will remain filled. In practice, this fills donut holes at this scale or smaller with minimal changes to the rest of the disk mask. This kernel size was chosen as it is larger than most holes that appear in the spiral structures of these galaxies. Larger kernel sizes or successive iterations of this binary closing will begin to fill in additional cells near the edges of the disk and are therefore avoided. This operation is shown in more detail in Appendix~\ref{sec:disk_definition_appx}.

We define the orientation of each system such that it is aligned with the angular momentum of the inner disk. To calculate this, we take the average angular momentum vector of the cold gas ($T < 10^{4}~\rm{K}$) within 7 kpc (physical) of the halo center. To separate the misaligned features (polar ring and warp) from the inner disk, we impose a cut both in radius and in angular momentum orientation. Gas with angular momentum vectors within 15\textdegree\ of the system's orientation vector is classified as inner-disk. To prevent gas in the central-most regions, which may have large velocities, from being misclassified, we additionally classify all gas within the inner co-moving 5 kpc as inner-disk.  Thus, gas outside the inner co-moving 5 kpc and with a greater than 15\textdegree\ angular momentum vector offset is classified as outer-disk/misaligned.
\begin{figure*}[ht!]
\includegraphics[width=\textwidth]{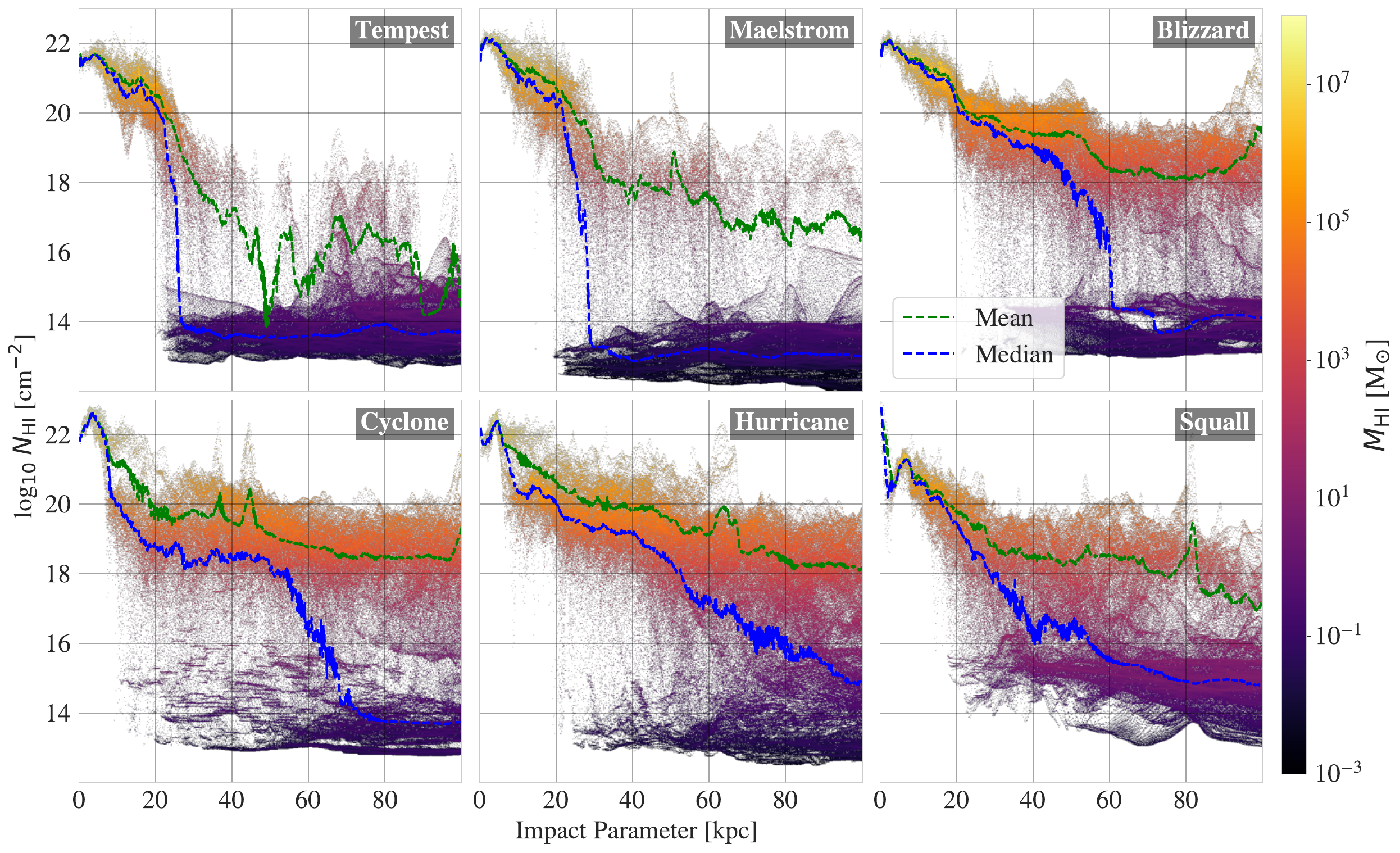}
\caption{
Radial profiles for face-on \HI\ column density maps at redshift \zzero~(see Fig.~\ref{fig:hi_projections}).
The heat map shows a 2-D histogram of pixels in the face-on projection, color-coded by the \HI\ mass of each pixel at a given point on the curve.
The green dashed line shows the mean profile, while the blue dashed line shows the median profile. The mean and median column densities are similar within the disk itself. Outside the disk, the median profile drops sharply, and the mean profile falls off more gradually. In the CGM, the mean values are dominated by the densest clumpy regions.
The top row shows the \LessActive\ systems, which drop off more sharply, while the bottom row shows the \MoreActive\ systems, which drop off more gradually.
}
\label{fig:hi_curves}
\end{figure*}

\section{Extended \HI\ Profiles} \label{sec:hi_profiles}
The six FOGGIE galaxies show distinct \HI\ profiles at redshift $z=0$, both in the extended \HI\ disks and in the CGM; however, some general trends remain consistent across all systems. We begin by \rev{classifying our systems based on their local environment}. 

\subsection{Classification of Halos Based on Local Environment}

There is a known relation between the presence of satellites and the abundance of \HI\ in the CGM, related both to merger events and the tidal stripping of satellite material \citep{hani18,saeedzadeh23,roy24}. Fig.~\ref{fig:classification_quantification} displays this correlation for the FOGGIE halos, showing the relative abundance of \HI\ in their CGMs versus the satellite mass within 2$R_{\rm vir}$, normalized to the main halo virial mass. The six halos considered in this study occupy a continuous distribution along this relation. \rev{At times, we will present} these systems in two categories based on their position on this relation: halos that are \LessActive\ with \HI\ and halos that are \MoreActive\ with \HI, \rev{however, we emphasize that they lie on a continuum.} Importantly, this environmental classification corresponds to various properties of the disk. This is evident in Fig.~\ref{fig:hi_projections}, which shows face-on and edge-on \HI\ column density projections for the six halos. \rev{In all subsequent figures, halos will be presented in order from \textit{Least populated} to \textit{Most populated}}

The \rev{three} \LessActive\ halos are comprised of those on the bottom left of the relation in Fig.~\ref{fig:classification_quantification}, including Tempest, Maelstrom, and Blizzard. These systems have relatively little dense material in their CGM, form thin, \rev{coherently rotating} gaseous disks, and do not show the presence of polar rings at $z=0$, although both Blizzard and Tempest had polar rings during their evolution. \rev{Note, we identify the formation of a thin gaseous disk (or vertical disk settling) as a decrease in aspect ratio (height over radius) at a given radius \citep{hopkins23-DiskTransitions} (see Sec.~\ref{sec:disk_evolution}).}

The three \MoreActive\ halos are comprised of those on the top right of the relation, including Cyclone, Hurricane, and Squall. These systems have more high column density material in their CGMs and \rev{have multiple distinctly misaligned components at $z=0$. While the innermost disks of these systems can still be relatively thin, they do not show clean disk settling, tend to be thicker than the \LessActive\ systems, particularly at higher radii} (see Table~\ref{tab:halo_masses}, Sec.~\ref{sec:disk_evolution}), and have significant misaligned features at $z=0$. Importantly, there is no correlation between \rev{a system's virial mass and its location along this continuum}. We note that Blizzard and Cyclone are edge cases of their respective classifications, with Cyclone showing significant evolution at late times. Hurricane, additionally, is the only system with a star-forming polar ring \citep{wright24}.  

\subsection{Description of Extended Disks} \label{sec:desc_of_extended_disks}

All six galaxies show spiral structures with characteristic \HI\ radii on the order of 20-50 kpc (Table~\ref{tab:halo_masses}, Fig.~\ref{fig:hi_projections}). Tempest and Maelstrom have the least populated CGMs at $z=0$, with a distinct boundary between the disk and CGM gas, and scattered clumps of \HI\ gas throughout the nearby CGM. When viewed edge-on, a relatively strong ($\sim 20$\textdegree) warp is visible in the outer disk of Tempest. Blizzard is a borderline case, with a more extended disk structure and more high column density \HI\ gas in the nearby CGM. This is largely due to environmental factors, with several interacting satellites visible in the projections. Similar to Tempest, Blizzard shows a strong warp in its extended disk and very similar overall disk morphology.

Cyclone is another borderline case. Despite having a smaller inner disk and a clearly visible polar ring, its disk is overall more ordered, which will be reflected in later quantifications. Squall and Hurricane have compact inner disks, misaligned from outer regions. They additionally have very diffuse extended \HI\ past the disk definition. In the case of Hurricane, a polar ring is also visible, implying three distinct misaligned components to this system (misaligned inner disk, outer disk, and polar ring).  The lifetimes of these features vary greatly. See Sec.~\ref{sec:disk_evolution} for a discussion of how these features arise and evolve.

The quantitative similarities and differences between these \HI\ profiles are emphasized in Fig.~\ref{fig:hi_curves}, which shows the median and mean column density profiles. In general, each system shows a sharp drop off in median \HI\ column density at the disk edge, dropping from values of $10^{20}$--$10^{22}~\rm cm^{-2}$ within the disk to values of $10^{14}$--$10^{16}~\rm cm^{-2}$ outside the disk. The three \LessActive~systems in the top row (particularly Tempest and Maelstrom) show sharper drop-offs than the more extended and interacting systems. The mean \HI\ column densities (green curve) behave similarly to the medians within the disk, but begin to differ in the extended disks and vary drastically within the CGM. This is due to the clumpy nature of the \HI\ gas within the CGM of the FOGGIE simulations \citep{Augustin25}. For the three \MoreActive\ systems in the bottom row (e.g., Cyclone, Hurricane, Squall), the mean column density drops much more gradually, not showing as sharp a transition as the other systems or the median curves.

Since these clumps comprise a relatively low covering fraction at higher impact parameters, the median values are insensitive to them. Other simulation studies \citep{mina21,trapp22,ramesh23,piacitelli25} see similar drop-offs in total \HI\ column density, but do not necessarily see spikes in the mean where small-scale overdensities dominate, likely due to limited CGM resolution. While observations will be biased towards detecting the upper envelope of these distributions, the difference between these mean and median profiles is significant in the interpretation of observations with a limited number of sightlines (e.g., absorption studies) and may limit the effects of stacking multiple \HI\ column densities at higher impact parameters \cite[e.g.,][]{veronese2025}.

\rev{The central-most \HI\ column densities in these systems are high, likely driven by the densely concentrated stellar profiles. Column densities are particularly high in the \MoreActive\ systems ($\rm{N}_{HI}>10^{22} \rm{cm}^{-2}$). These systems have denser stellar profiles in the innermost region \citep{wright24} and more extreme metallicity gradients in the innermost 5 kpc \citep{Acharyya24}. They also have misaligned features that are projected at high inclinations onto the central pixels, increasing column densities. This is most significant for Squall, in which our chosen orientation results in the compact inner component being projected at high inclination.}

Taken together, Fig.~\ref{fig:hi_projections} and Fig.~\ref{fig:hi_curves} paint the picture of extended \HI\ disks with a distinct boundary in \HI\ column density, surrounded in the CGM by small-scale clumps that dominate the mean column density but have low covering fractions, and in some cases diffuse extended material associated with interacting satellites and misaligned components. We are able to resolve the high column density clumps in the CGM due to the forced spatial resolution within the FOGGIE simulations. Similar extended \HI\ structures can be seen in other simulations that emphasize spatial resolution in the CGM \citep{nelson16,vandevoort19,hummels19,ramesh23}.

This specific structure of the \HI\ gas presented here has significant consequences on the observability (both the small spatial scales and low covering fractions of the clumpy material and diffuse extended material) of this material. Specific consequences for observational resolution effects and interferometry are discussed in \citetalias{trapp25b}. \rev{Similar to what is seen in other recent simulation comparison studies \cite[e.g.][]{marasco25:mhongoose-sim-comp,lin2025-FeastsCompWithSimulations}, \MoreActive\ systems tend to have more low column density \HI\ than is seen in observed galaxies.}

\begin{figure}[ht!]
\includegraphics[width=0.48\textwidth]{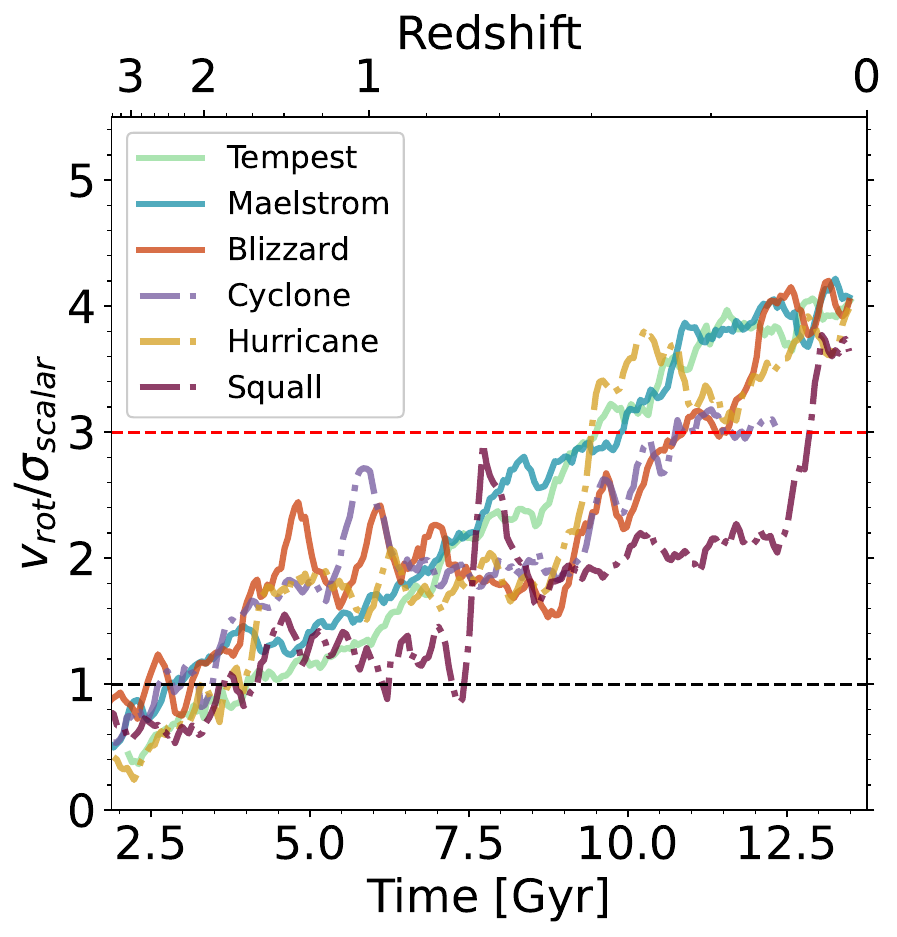}
\caption{
Kinematic order of the six disks in this study. Curves show density-weighted mean ratios of rotational velocity and velocity dispersion ($v_{rot}/\sigma$) within our disk definition. The lower dashed line (black) denotes the upper limit for a dispersion-dominated system ($v_{rot}/\sigma<1$) while the upper dashed line (red) provides a more stringent cutoff ($v_{rot}/\sigma>3$). All systems quickly become rotationally supported and continue to settle regardless of classification (denoted by line type). Curves were smoothed by a moving average filter (270 Myr) to improve readability. 
\label{fig:disorder_evolution}}
\end{figure}

\begin{figure*}[ht!]
\includegraphics[width=\textwidth]{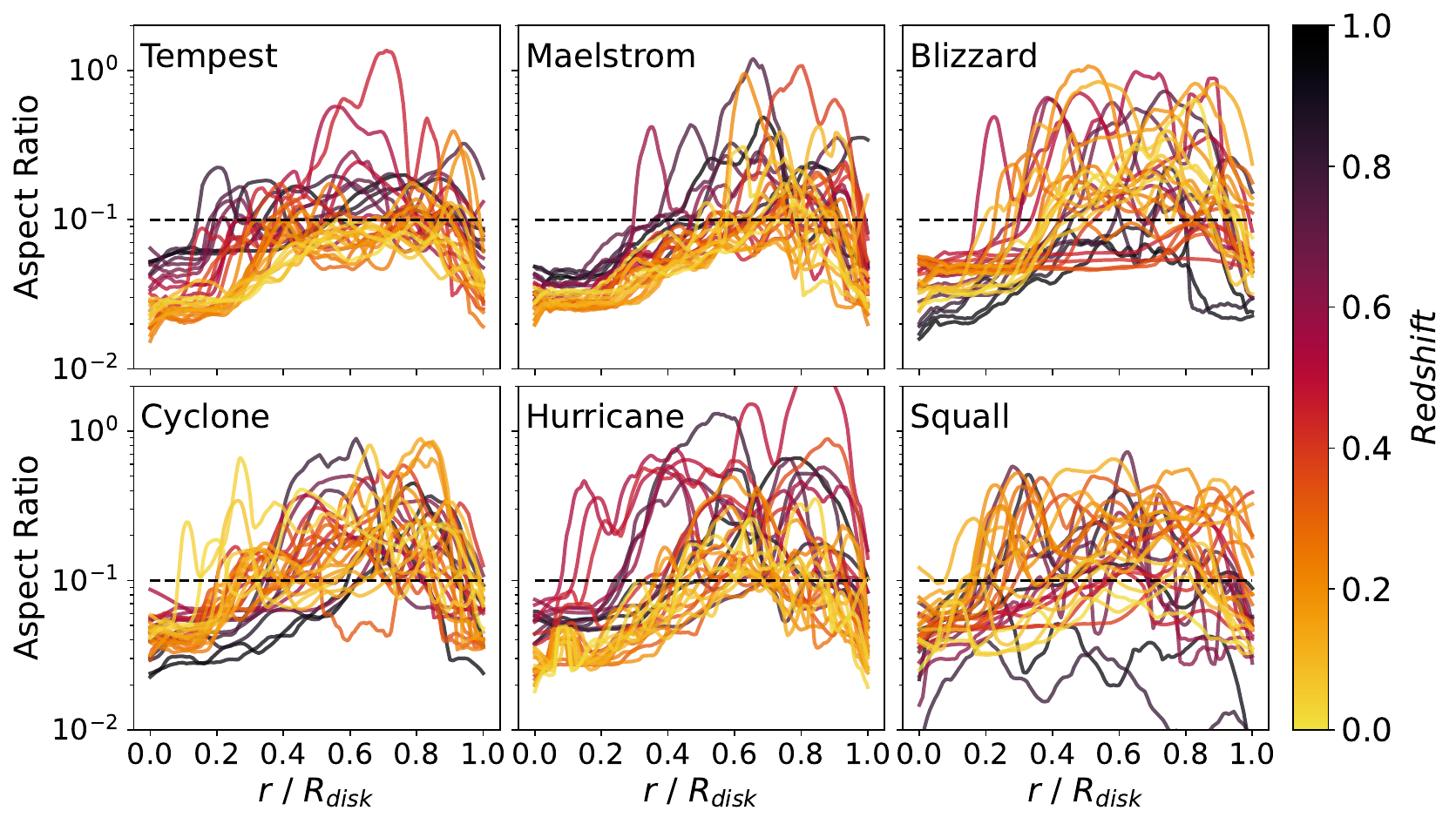}
\caption{
\rev{Disk aspect ratio (\HI\ half-mass height over half-mass radius) as a function of galactocentric radius. Smaller aspect ratios correspond to thinner disks. Different colored lines show evolution with redshift. The \LessActive~systems (top) show gradual vertical disk settling and form thinner, extended disks, with very thin central disks at $z=0$. The \MoreActive~systems (bottom) go through periods of thickening and tend to be thicker, although the inner regions can still be relatively thin.}
\label{fig:thin_disks}}
\end{figure*}

\section{Time Evolution of \HI\ Disks} \label{sec:disk_evolution}

In this section, we show how the \HI\ disks in all six systems evolve over cosmological time between redshifts $z\sim3.5$ and $z\sim0$. We first discuss qualitatively how each system evolves. We then quantify how the systems settle, heat their inner CGMs to near the virial temperature, and form thin disks. The systems \rev{show differences in some of these evolutionary stages based on how relatively populated their inner CGMs are.} 

\subsection{Qualitative Description}

All systems develop some form of a polar ring or warp at some point in their history. The formation of these misalignments is often preceded by or concurrent with an interaction/merger with a satellite galaxy; however, the presence, strength, and lifetime of these structures are highly system-dependent. A deeper analysis of the evolution of individual systems will be presented in Sec.~\ref{sec:tempest_blizzard_comparison} for Tempest and Blizzard, and in Appendix~\ref{sec:other_halo_time_evolutions} for the other halos. Here, we give a brief summary of the evolution of all systems.

The \rev{three} \LessActive~systems have various misalignments during their evolution, but do not have polar rings at $z=0$. Tempest forms a polar ring $\sim$1 Gyr after a merger event, which persists for $\sim$4 Gyr before precessing to form a warp. Maelstrom excites a few warps, but these decay relatively quickly ($\sim$1 Gyr), and the system never forms any polar rings. Blizzard forms polar rings at two different times. The first forms as a subhalo approaches, and is disrupted during the resultant merger event after $\sim$1 Gyr. The latter forms similarly with another approaching subhalo, but is disrupted after $\sim$1 Gyr. Following this, a warp develops.

The \MoreActive~systems all show significant misalignments at $z=0$ (polar rings/misaligned inner disks). Squall forms a polar ring, and the inner disk is gradually destroyed over the next $\sim$4 Gyr. By $z=0$, the inner disk has shrunk to a radius of $\sim$2 kpc, while the polar ring extends out to $\sim$30 kpc. Hurricane forms multiple, concurrent polar rings around $t=11$ Gyr ($z\sim0.2$), which persist until $z=0$. Cyclone forms a large polar ring and has a relatively small inner disk by $z\sim 0$.

\rev{All systems show a similar evolution in disk size and gas mass, which is discussed in more detail in Appx.~\ref{sec:disk_size_evolution}.} A detailed comparison of the HI Size-Mass relation \citep{wang16-HISizeMass} with radio observations is presented in \citetalias{trapp25b}.

\begin{figure*}[ht!]
\includegraphics[width=\textwidth]{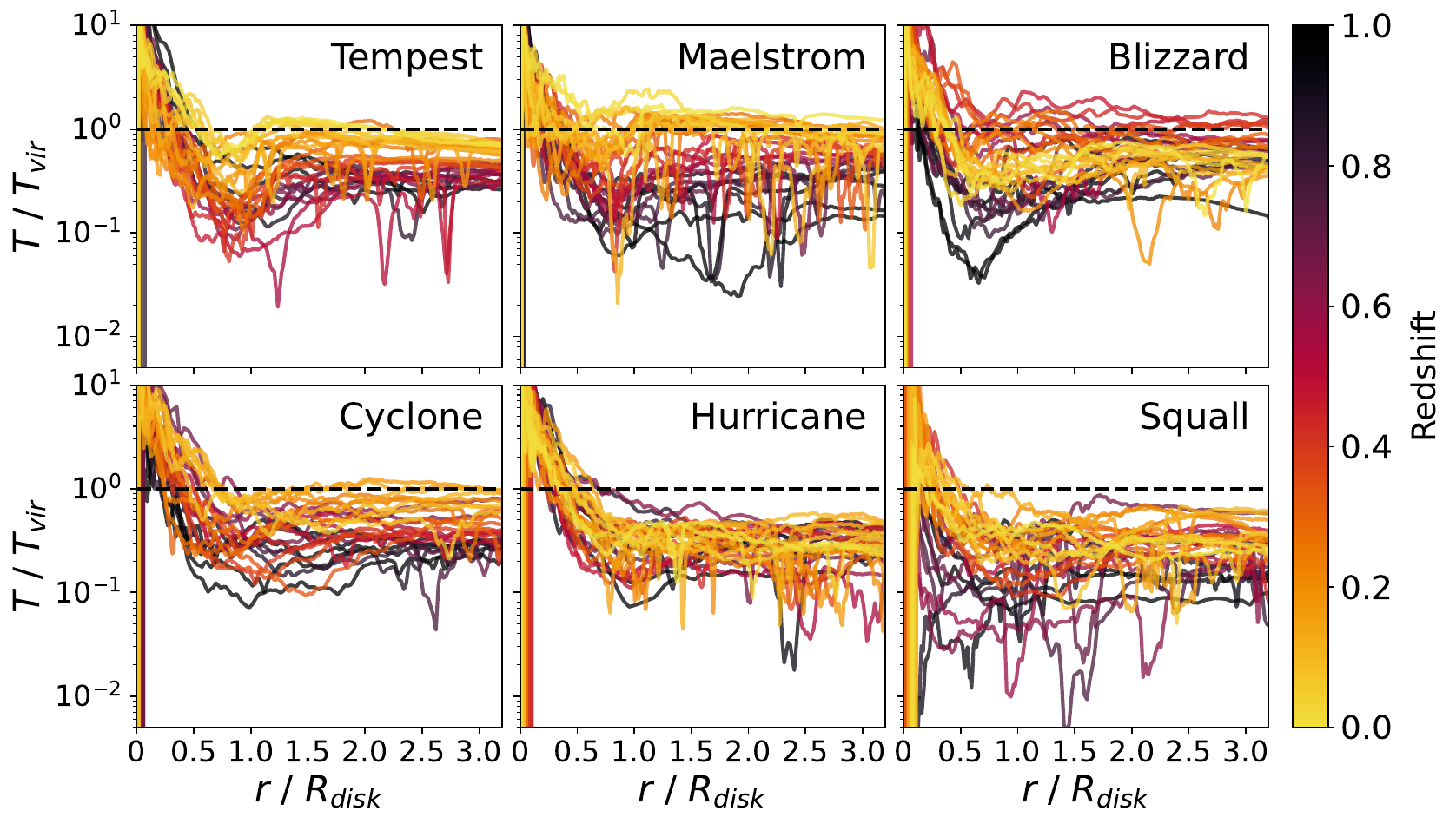}
\caption{
Mass-weighted mean temperature of the CGM normalized to the halo virial temperature $T_{\rm vir}$ as a function of galactocentric radius, normalized by the disk radius ($R_{\rm disk}$ corresponds to $\sim$0.1--0.15 $R_{200}$, Table~\ref{tab:halo_masses}). The three \LessActive~systems (top) generally show a hotter inner CGM, while the \MoreActive~systems (bottom) have cooler inner CGMs. Different colored curves show the evolution from redshift $z=1$ to $z=0$, while the dashed line shows the virial temperature.  Blizzard's CGM temperature drops below the virial temperature at late times, which is likely related to the flaring in the disk. On the other hand, the inner CGM of Cyclone is just beginning to warm up, implying it may settle soon.
\label{fig:iCGM_virialization}}
\end{figure*}

\subsection{Disk Settling} \label{sec:disk_settling}

Fig.~\ref{fig:disorder_evolution} shows the evolution of kinematic ordering ($v_{rot}/\sigma$) for the six disks, where $v_{rot}$ is the total rotational velocity and $\sigma$ is the velocity dispersion. Velocity dispersion was calculated as a combination of the bulk scalar dispersion of a cell within a 1.5 kpc kernel and the cell's inherent thermal velocity dispersion. As expected, kinematic ordering more or less increases monotonically with time as disks settle \citep{simons17}. All systems evolve similarly at early times, becoming rotationally dominated ($v_{rot}/\sigma>1$) by $t\sim4$ Gyr ($z\sim1.6$). Tempest and Maelstrom, the two least populated systems, increase in kinematic order more consistently, reaching the stronger cutoff ($v_{rot}/\sigma>3$, \citealt{simons17}) by $t\sim10$ Gyr ($z\sim0.3$). Blizzard, Hurricane, and Cyclone have periods where order does not increase, followed by periods with more rapid increases. This is likely due to their more active interaction/merger histories.  Squall is the major outlier, showing a rapid increase in order followed by a rapid decrease during the 2:1 merger at $t\sim7.5$ Gyr ($z\sim0.7$). From here, the order remains relatively constant until near $z=0$, where it increases sharply. \rev{While overall evolution is similar, the \MoreActive\ systems tend to lie systematically lower at late times, with Hurricane showing the most rotational support.}

\rev{Fig.~\ref{fig:thin_disks} shows the aspect ratios (half mass height over half mass radius) of all six \rev{gaseous} disks as a function of radius and redshift. Half-mass height was measured along the direction of the average angular momentum at each radial bin. We identify thin disk formation (or vertical disk settling) as the decrease in this aspect ratio at a given radius \citep{hopkins23-DiskTransitions}.}

\rev{The \LessActive~systems tend to show more classical vertical disk settling, while the \MoreActive~systems show thicker disks on average with more chaotic evolutions. Tempest and Maelstrom, the \textit{Least populated}~systems show a gradual decrease in aspect ratio at all radii from $z=1$ to $z=0$, with flaring in the outer regions. Blizzard shows a thin disk at $z=0$, which thickens following subhalo interactions. Following this, the inner region vertically settles, while the outer region becomes increasingly flared.}

\rev{Similar to Blizzard, The \MoreActive~ systems tend to show more chaotic evolution and tend to be thicker on average. Cyclone's inner disk goes through periods of thickening and thinning, and shows higher aspect ratios at larger radii. Hurricane shows very high aspect ratios throughout its evolution. However, by $z=0$, it has a relatively thin inner disk, as reflected by its high rotational support seen in Fig.~\ref{fig:disorder_evolution}. Squall tends to have a thicker disk at all radii that varies strongly with redshift.}


Fig.~\ref{fig:iCGM_virialization} shows the corresponding mass-weighted mean CGM temperatures for these systems, normalized to the virial temperature. Generally speaking, the three \LessActive~systems show hotter inner CGMs close to their respective virial temperatures by $z=0$, while the three \MoreActive~systems have cooler inner CGMs. The mean temperature of the inner CGM is expected to be lower in the \MoreActive~systems, since they have more cold material in their CGM, by definition. \citet{Stern21} found, in a different set of simulations, that there is a correlation between inner CGM temperature approaching the virial temperature and the formation of thin disks, which they refer to as ``inner CGM virialization." Here, we do not claim that inner CGM temperatures near the virial temperature imply that the galaxies have virialized their inner CGM --- \citet{lochhaas21} found that the FOGGIE galaxies are far out of virial equilibrium in the inner CGM near $z=0$ because significant non-thermal gas motions are driving the total energy content of the system to large values that are not balanced by the gravitational potential energy. However, there appears to be a correlation between whether a system is \LessActive~or \MoreActive~ and the thickness of its disk, but not with halo or disk mass, implying local environmental effects including CGM gas mass and temperature have influences on thin disk formation.

We discuss two edge cases within each category: Blizzard and Cyclone. Near $z=0$, Blizzard's inner CGM drops below the virial temperature. This is likely due to the destruction of the polar ring and subsequent pollution of the CGM with cold dense gas (see Sec.~\ref{sec:blizzard_time_evolution}). This is reflected in Fig.~\ref{fig:thin_disks}, which shows Blizzard's disk beginning to flare at later times.
Contrarily, Cyclone's inner CGM is just approaching its virial temperature near $z=0$, so it is likely to continue to settle in the future as it continues to evolve along this continuum.

\section{Comparison of the origin and fate of polar rings in two systems} \label{sec:tempest_blizzard_comparison}

\begin{figure*}[ht!]
\begin{centering}
\includegraphics[width=\textwidth]{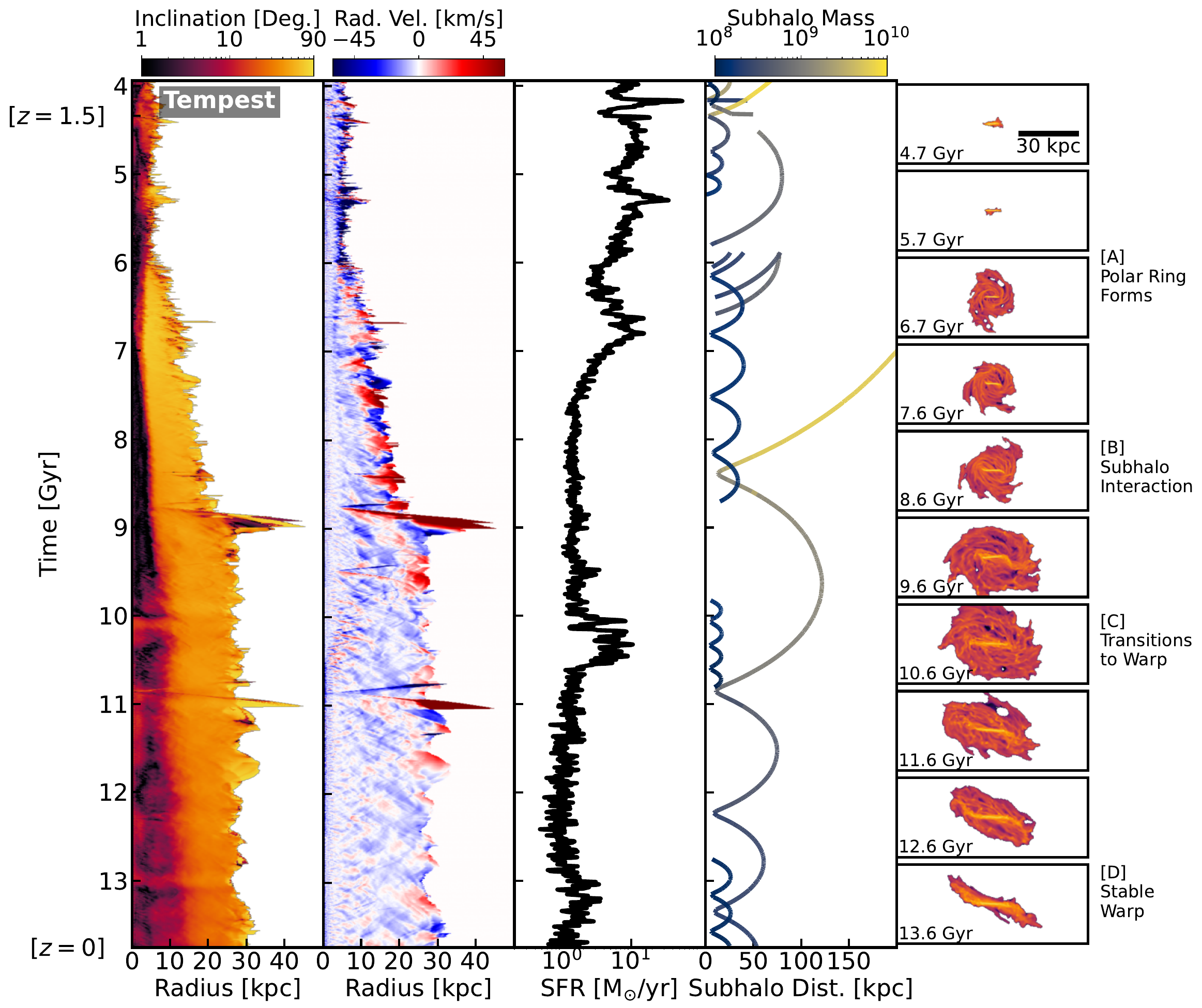}
\caption{
Several key properties of Tempest as a function of time, highlighting the precession of the polar ring formed at $t\sim6$ Gyr to a warped outer disk. The first two panels show ring-averaged values for key properties of the disk as identified by our disk selection criteria. Inclination (\textit{left}) shows the average tilt of the gas with respect to the disk plane. Nonzero values correspond to some degree of misalignment or warping. The radial velocities (\textit{left-middle}, positive/red is away from galactic center) show distinct spatial and temporal variance within the polar ring, warp, and inner disk. The large spikes seen in the inclination and radial velocity plots (t$\sim$8.5 and t$\sim$11 Gyr) are the result of subhalo interactions. The third panel (SFR) shows the total star formation rate of the system, which transitions from bursty to more stable at $t\sim7$ Gyr. The fourth panel shows satellite trajectories, color-coded by subhalo mass. Gaps in satellite trajectories appear when {\sc ROCKSTAR} cannot reliably identify the subhalo, which occurs most frequently during close passages between galaxies. The images on the right show edge-on density projections of the gas at representative points during this evolution. The text on the right (A-D) highlights key timepoints during the evolution of this system, which will be referred to in future figures.
\label{fig:tempest_time_plots}}
\end{centering}
\end{figure*}

\begin{figure}[ht!]
\includegraphics[width=0.49\textwidth]{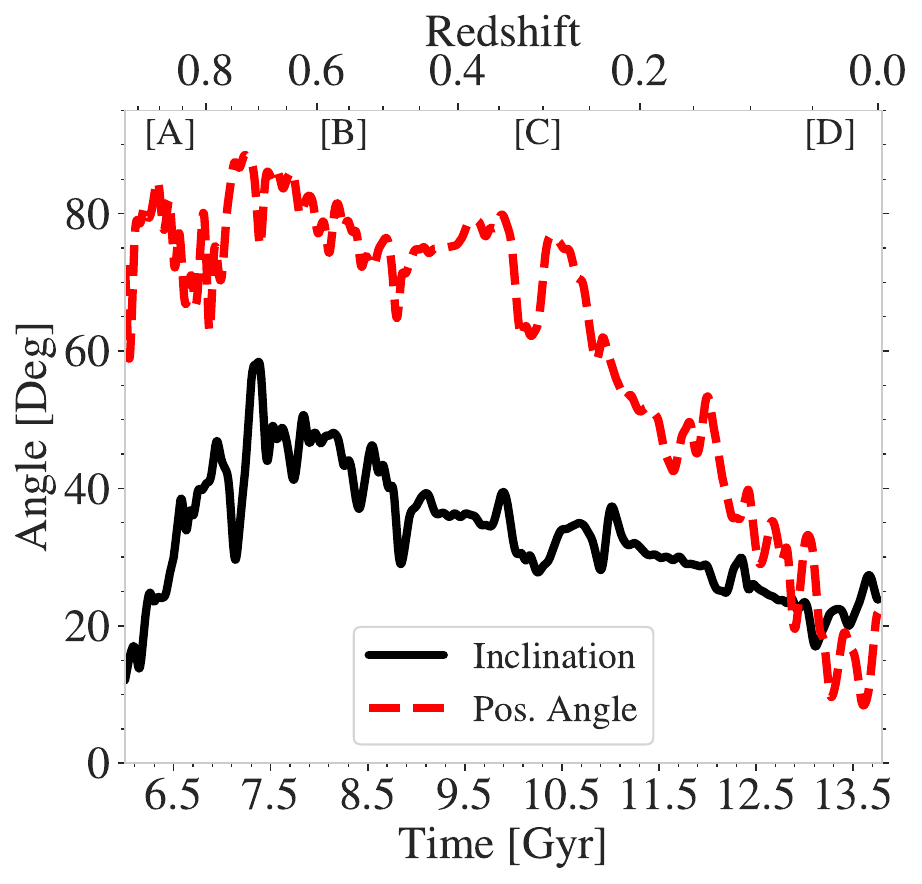}
\caption{
The precession of the polar ring in Tempest into a warp. The black-solid and red-dashed lines show the mean inclination and position angle of the polar ring, respectively. Letters (A-D) correspond to events shown in Fig.~\ref{fig:tempest_time_plots}. From t$\sim$7--10 Gyr, the polar ring gradually precesses towards the disk. Around t$\sim$10 Gyr, a warp begins to form within the inner disk $\sim$90\textdegree\ from the orientation of the polar ring. Following this, the polar ring continues to precess towards the disk plane, while simultaneously rotating in position angle to orient itself with the warp in the inner disk.
\label{fig:tempest_ring_evolution}}
\end{figure}

\begin{figure}[ht!]
\includegraphics[width=0.49\textwidth]{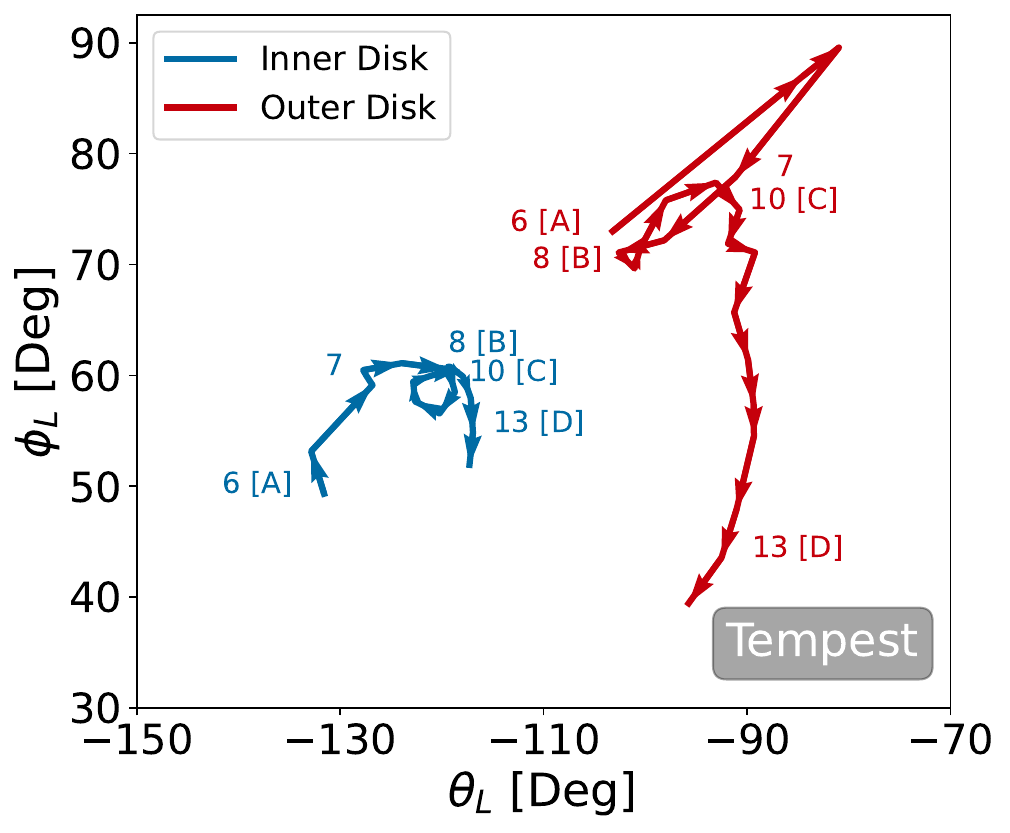}
\caption{
Orientation evolution of the angular momentum vector of the aligned (inner, blue) and misaligned (outer, red) disks in Tempest between $t= 6-13.7$ Gyr. Text values show the times in Gyr and corresponding events (A-D) shown in Fig.~\ref{fig:tempest_time_plots}. The inner disk does not precess significantly. The outer disk's orientation evolves more significantly and is clearly distinct from the inner disk.  Each point is the average value over 350 Myr. 
\label{fig:tempest_orientation_evolution}}
\end{figure}

\begin{figure*}[ht!]
\includegraphics[width=\textwidth]{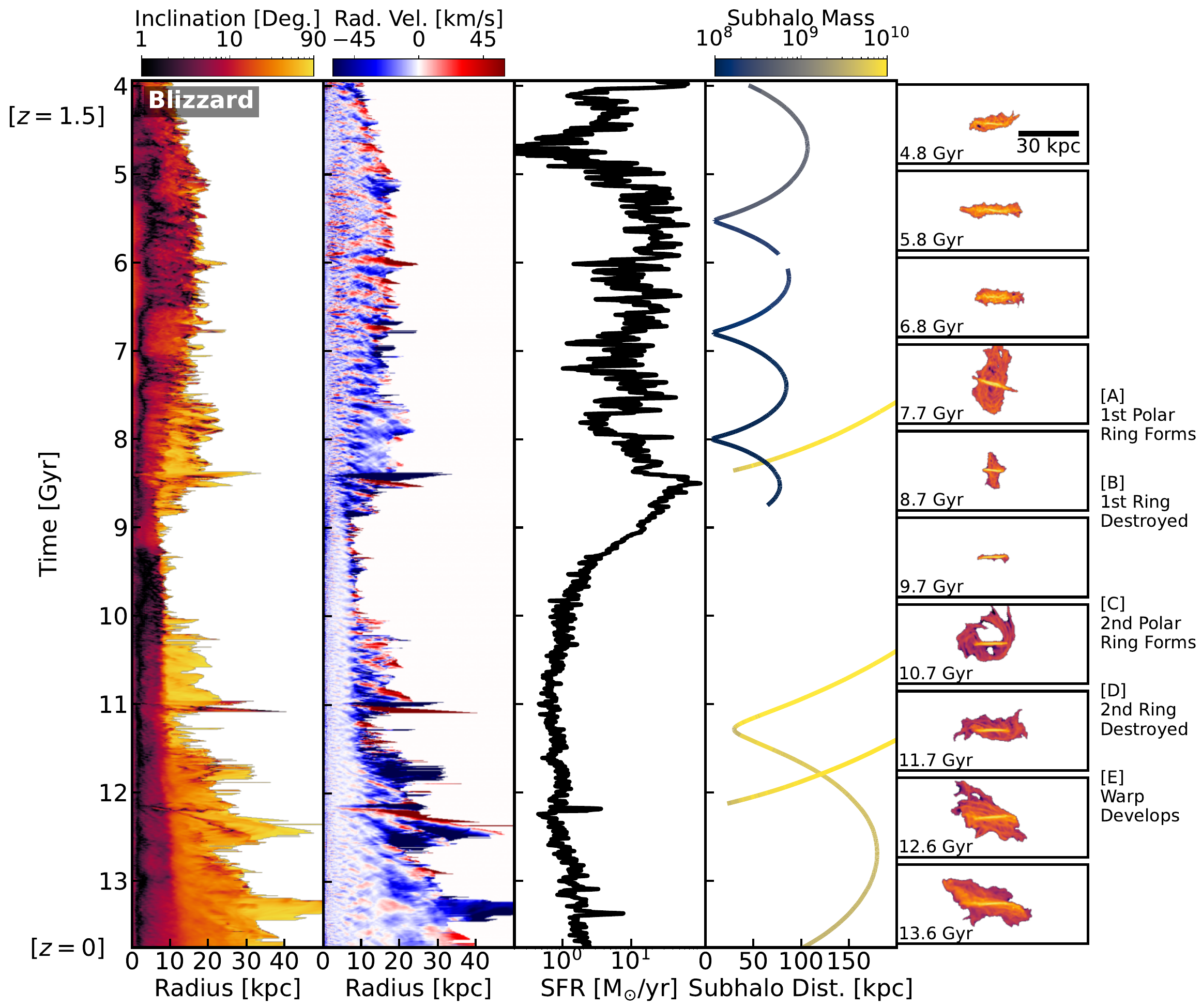}
\caption{
Same as Fig.~\ref{fig:tempest_time_plots}, but for Blizzard. Two polar rings form at t$\sim$7.5 and t$\sim$10 Gyr. Unlike Tempest, these polar rings are relatively short-lived. Both are quickly dissipated ($\sim1$ Gyr), likely by the merger events. The dissipation of this first polar ring is accompanied by a decrease in star formation, which then continues to rise as the warp rapidly feeds the inner disk.
\label{fig:blizzard_time_plots}}
\end{figure*}

\begin{figure}[ht!]
\includegraphics[width=0.49\textwidth]{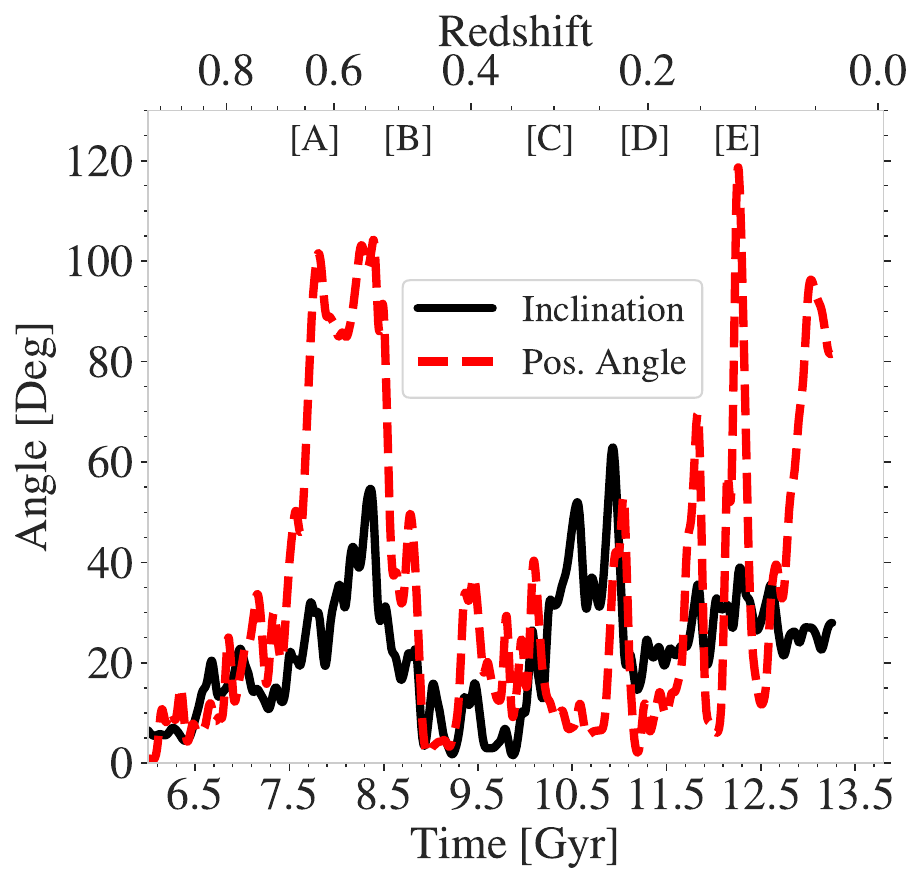}
\caption{
Same as Fig.~\ref{fig:tempest_ring_evolution} but for Blizzard. The inclination peaks when a polar ring is present; however, unlike in Tempest, these features are much less stable and are disrupted on timescales of $\sim$1 Gyr.
\label{fig:blizzard_ring_evolution}}
\end{figure}

\begin{figure}[ht!]
\includegraphics[width=0.49\textwidth]{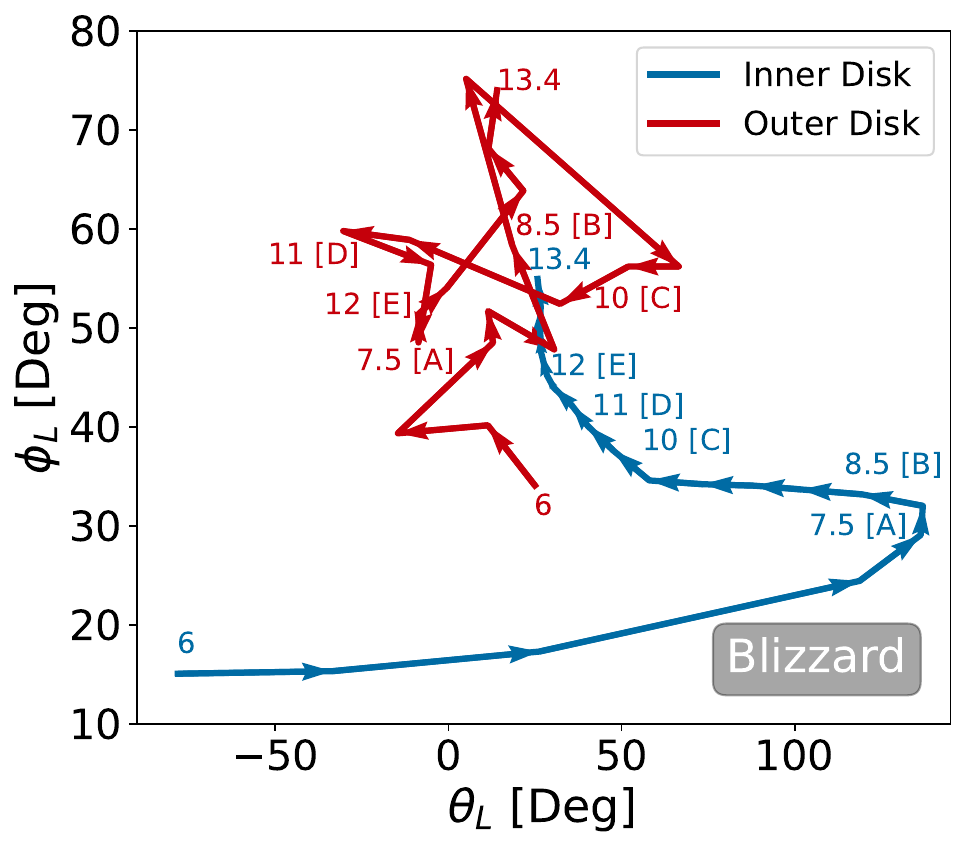}
\caption{
Same as Fig.~\ref{fig:tempest_orientation_evolution}, but for Blizzard. The orientation of all components varies more significantly (note the difference in axes). The inner disk precesses significantly. The outer disk's evolution is not as smooth as the various misaligned components have independent origins. Letters (A-E) correspond to events shown in Fig.~\ref{fig:blizzard_time_plots}. 
\label{fig:blizzard_orientation_evolution}}
\end{figure}


\rev{Despite both developing polar rings and warps throughout their evolution and having strikingly similar disk structure at $z=0$, Tempest and Blizzard undergo significantly different evolutions. In this section, we investigate how the polar ring formation, evolution, lifetimes, and warp formation mechanisms are drastically different between these two systems. We generally find that the late-time evolution of Blizzard, wherein its inner CGM is cooler, and its gaseous disk thickens, may be a consequence of the destruction of its polar ring and subsequent pollution of its CGM with dense material. The polar ring in Tempest, on the other hand, precesses gradually to form a warp at late times, and the system has a relatively unpopulated CGM.}

In the following subsections, we consider the radial dependence of several key properties, including inclination and radial velocities. Ring-averaged values were calculated as the density-weighted mean value for a given parameter within a 0.274 kpc thick annulus (one cell at $z=0$). These averaged values only include gas cells within our disk definition (Sec.~\ref{sec:hi_disk_definition}). Inclination was calculated as follows:
\begin{equation}
    I = \tan^{-1}\bigg{(}\frac{z}{s \cos(\theta+PA)}\bigg{)}~,
\end{equation}
where $z$ is the height above/below the disk, $s$ is the cylindrical radius, $\theta$ is the azimuthal angle on the disk, and $PA$ is the average position angle of the ring of gas. Radial velocities in these plots correspond to spherical radial velocities towards (negative) and away from (positive) the center of the disk. Velocities were centered on the disk using the bulk velocity of stars and dark matter particles within the inner 3 kpc.

\subsection{The Precessing Polar Ring of Tempest}\label{sec:tempest_time_evolution}

At $z=0$, Tempest shows a large warp (up to 20\textdegree) in its extended disk. This warp forms as a polar ring following a merger event at $t\sim4$ Gyr ($z\sim1.6$). The polar ring then precesses towards the disk plane, eventually aligning itself with the inner disk and transitioning into a warp. We identify four time points which we will refer back to:

\begin{itemize}
    \item $[A]$ 6 Gyr ($z=1$): The polar ring begins to form
    \item $[B]$ 8 Gyr ($z=0.6$): Massive subahalo interaction
    \item $[C]$ 10 Gyr ($z=0.3$): Polar ring becomes a warp
    \item $[D]$ 13 Gyr ($z=0.06$): A stable warp is present
\end{itemize}

Fig.~\ref{fig:tempest_time_plots} shows the evolution of this system in terms of ring-averaged inclination, ring-averaged radial velocity, star formation rate, and subhalo merger/interaction history. The ring-averaged inclinations show the average tilt of the gas with respect to the disk plane. Non-zero values correspond to some degree of misalignment or warping. The polar ring initially forms at $t\sim6$ Gyr [A]. This feature originates following \rev{a 14:1} merger event with a $\sim10^{10}$ M$_{\odot}$ subhalo. Between $t\sim6$ Gyr and $t\sim10$ Gyr, this polar ring remains a distinct subcomponent of the galaxy, indicated by the sharp transition in the inclination plot. At $t\sim$10 Gyr [C], this transition becomes more continuous, indicating the polar ring has become a warp.
The radial velocities show distinct spatial and temporal variance within the polar ring, warp, and inner disk. Both the spatial and temporal scales of the radial velocity structure in the outer warped disk and polar disk are larger than in the inner disk. The warped disk is biased towards inward radial motion, implying it is feeding the inner disk. The star formation rate for this system transitions from bursty to more stable rates at t$\sim7$ Gyr.\footnote{We distinguish `bursty' and `stable' SFRs largely qualitatively; however, there is a distinct change in mean SFR and its temporal variation for both Tempest and Blizzard. Tempest goes from SFRs of $5.4\pm2.1$ M$_{\odot}$ yr$^{-1}$ to $1.5\pm0.2$ M$_{\odot}$ yr$^{-1}$ in the Gyr before and after this transition, respectively. Similarly, Blizzard goes from SFRs of $9.6\pm7.9$ M$_{\odot}$ yr$^{-1}$ to $1.2\pm0.4~\rm{M}_{\odot}~\rm{yr}^{-1}$.}
The projections on the right show the density of the gas at various times along the same line of sight. Importantly, at $t=9.7$ Gyr, a small warp in the inner disk can be seen, which the polar ring aligns with by $z=0$.

The precession of the polar ring is quantified in Fig.~\ref{fig:tempest_ring_evolution}, which shows the ring-averaged inclination and position angle. Between $t\sim6.5$--$9$ Gyr, the polar ring precesses relatively rapidly towards the disk plane. At $t\sim10$ Gyr [C], the position angle of the misaligned gas begins to change rapidly, as it aligns with the smaller warp present in the inner disk.


Fig.~\ref{fig:tempest_orientation_evolution} shows how the orientation of these two vectors evolves between $z=1$ to $z=0$. The variables $\theta_L$ and $\phi_{L}$, first shown in \citet{simons25}, represent the orientation of the angular momentum vector and are defined as follows:
\begin{equation}
    \theta_{L}=\arctan\bigg{(}\frac{L_{x}}{L_{y}}\bigg{)},~ \rm{and}
\end{equation}
\begin{equation}
    \phi_{L}=\arccos\bigg{(}\frac{L_{z}}{L}\bigg{)}, 
\end{equation}
where $L$ is the average angular momentum of each component. The $x$, $y$, and $z$ components in this space follow the simulation grid, and are thus arbitrary with respect to the system. The orientation of the inner disk does not evolve significantly during this time period, remaining within 10\textdegree--20\textdegree. The orientation of the outer disk evolves more significantly, moving away from the inner disk during the initial formation of the polar ring, before precessing towards the disk plane.

\subsection{The Fragile Polar Rings of Blizzard}\label{sec:blizzard_time_evolution}

Similar to Tempest, Blizzard shows a large warp (up to 20\textdegree) in its extended disk at $z=0$. Unlike Tempest, which forms this warp through the precession of a long-lived polar ring, Blizzard forms two much shorter-lived polar rings and then separately develops a warp. We identify five time points which we will refer back to:

\begin{itemize}
    \item $[A]$ 7.5 Gyr ($z=0.7$): The 1st polar ring forms
    \item $[B]$ 8.5 Gyr ($z=0.5$): The 1st ring is destroyed
    \item $[C]$ 10 Gyr ($z=0.3$): The 2nd polar ring forms
    \item $[D]$ 11 Gyr ($z=0.2$): The 2nd ring is destroyed
    \item $[E]$  12 Gyr ($z=0.1$): The warp develops
\end{itemize}

Fig.~\ref{fig:blizzard_time_plots} shows the evolution of this system in terms of ring-averaged inclination, ring-averaged radial velocity, star formation rate, and subhalo merger/interaction history. Blizzard initially forms a polar ring at $t\sim7.5$ Gyr ($z\sim0.7$) [A]. Given that the formation of this precedes a \rev{66:1 merger} with a massive ($\sim10^{10} M_{\odot}$) subhalo, it likely originates from the tidal accretion of misaligned material. This polar ring is short-lived ($\sim1$ Gyr), however, and is destroyed when the subhalo merges with the main halo [B]. Following this, another polar ring is excited at $t\sim10$ Gyr [C], likely through a similar process with a different subhalo \rev{(50:1, $\sim10^{10} M_{\odot}$)}. This polar ring is also disrupted after $\sim1$ Gyr [D]. After this, a warp forms [E] and lasts until $z=0$. Similar to Tempest, the radial velocities of the misaligned components have larger spatial and temporal scales, and are biased towards inflow. Blizzard's star formation rate transitions from bursty to more stable at a later time ($t\sim8.5$ Gyr, $z\sim0.5$), coincident with the destruction of its first polar ring.

Fig.~\ref{fig:blizzard_ring_evolution}, which shows the inclination and position angle of these misaligned features, makes it clear that the orientations of these features are largely uncorrelated, implying they have different origins despite similar formation processes. The first polar ring has a drastically different position angle than the second, despite having a similar inclination. Similarly, the formation of the warp is accompanied by a sudden shift in the inclination and position angle, as opposed to the smooth precession seen in Tempest.

The differences between Blizzard and Tempest are further emphasized in Fig.~\ref{fig:blizzard_orientation_evolution}, which shows the orientation of the angular momentum vectors of Blizzard's inner and outer disks. There are several important differences between this plot and Fig.~\ref{fig:tempest_orientation_evolution}. Most obvious is the fact that the inner disk's orientation precesses significantly, especially at early times. Eventually, the orientation precesses towards the outer disk at late times. The orientation of the outer disk also evolves sporadically, and does not show continuous precession as is seen in the inner disk and in Tempest's components. This is consistent with the picture of various misaligned components being formed and destroyed.

\section{Discussion} \label{sec:discussion}

\subsection{What causes thin disk formation?}

 \rev{In this study, we place our six halos on a continuum based on the relative amount of \HI\ in their CGMs.}
The \LessActive~systems have a relatively small amount of dense CGM material, while the \MoreActive~systems have more. This classification is based on local environmental factors, including the content of the CGM and the total mass of satellites (Fig.~\ref{fig:classification_quantification}). Additionally, the \rev{location of a system along this continuum} correlates with several key properties of the disk itself. 
The \LessActive~systems all form thin, coherently rotating disks by $z=0$, but the \MoreActive~systems do not. 
The virial mass and disk mass of these systems are uncorrelated with this categorization, and all systems kinematically settle in a similar fashion. These categorizations also have a significant impact on the observable parameters of these systems, and are discussed further in \citetalias{trapp25b}.

These findings raise an important question. First, why do the \MoreActive~systems have cooler inner CGM gas? These systems are all more massive than the smallest \LessActive~system (Tempest), and two of them are more massive than all of the other systems.
\rev{Halos of this mass  ($\sim10^{12}$ \msun) at $z=0$ are expected to be near or past the transition to  ``hot mode" accretion \citep{keres05,Hafen22}.}
Other simulation studies \cite[e.g.,][]{chan22} have shown that non-thermal support terms can have drastic impacts on CGM temperatures and morphology. As shown previously in the FOGGIE simulations, non-thermal support terms such as turbulent pressure and bulk flows lead to cooler CGM temperatures in these systems; however, the halo is still virialized at the virial radius at $z=0$ \citep{lochhaas21}.

\rev{\citet{roy24} showed in a suite of idealized simulations that satellites of Milky Way mass galaxies seed the CGM with cool gas, with systems without satellites showing three orders of magnitude less cold gas. In their simulations, satellites seeded cold gas through direct tidal stripping of gas and induced cooling in the mixing layer of this stripped gas.} Therefore, we posit that the reason these massive systems do not show a hot inner CGM at these redshifts is largely an environmental effect. The massive, gas-rich satellites likely end up \rev{seeding the CGM with cold gas via tidal stripping and induced cooling} as they interact and merge with their host halos.  The temperature of this clumpy material will be lower on average than the virial temperature of the halo, and also not necessarily aligned with the rotation of the inner disk. \rev{For more information on tidal stripping and clump structure in these systems, see \citet{simons20} and \citet{Augustin25}, respectively}. This source of cool gas drives mean inner CGM temperatures down, and its kinematic misalignment with the inner disk inhibits \rev{the formation of thin, extended disks}. These results suggest deep and previously unexplored connections between CGM gas and inner disk properties. 

This is consistent with the picture of the angular momentum of these systems presented in \citet{simons25}, which shows that the angular momentum of the central galaxies does not begin to stabilize until $z\sim1$, with orientations largely determined by the combined angular momentum of cold gas accretion and mergers. It is only after there is a relatively small amount of misaligned, cold gas in the inner CGM compared with the virial mass that the angular momentum of the central galaxy can stabilize. This is clearly apparent in Tempest (Fig.~\ref{fig:tempest_orientation_evolution}), wherein the inner disk's orientation does not change significantly. In Blizzard (Fig.~\ref{fig:blizzard_orientation_evolution}), the orientation of the inner disk precesses much more significantly, but this slows down at $t\sim8$ Gyr ($z\sim0.6$)

These findings lead to several other questions. If these local environmental effects are the dominant factor preventing thin disk formation in the \MoreActive~systems, what are the timescales on which this clumpy material will either be destroyed or accrete onto the central galaxy? If most of it accretes, what implications does that have for the morphology of the galaxy? As shown in \citet{Augustin25}, at $z=1$ most clumps in these systems are long-lived ($t_{\rm cool}<t_{\rm shear}$). If the local environment continues to be highly populated (e.g., the incredibly active merger history of Hurricane seen in Appendix~\ref{sec:hurricane_time_evolution}), will the system eventually become massive enough that these clumps will be more rapidly heated up and destroyed?

\subsection{Key differences between Tempest and Blizzard}

Tempest and Blizzard both form thin \rev{gaseous} disks, albeit with strong warps in the outer disk. As emphasized in Fig.~\ref{fig:hi_projections}, the inner and outer disks of these systems look remarkably similar; however, their CGMs are drastically different. When analyzing the formation and evolution of the warps in these systems, we similarly see two different formation mechanisms and histories that ultimately lead to similar morphology. Tempest forms a long-lasting, stable polar ring at $t\sim6$ Gyr arising from a merger event that eventually precesses to comprise the warp at $z=0$. Blizzard, on the other hand, forms two polar rings associated with interacting subhalos, which are destroyed during the closest interaction with said subhalos. The warp then forms distinctly from the polar rings.

Similar to the differences between the \LessActive~and \MoreActive~systems, we argue that the differences in origin of these morphologies are largely environmental. Tempest is less massive than Blizzard and has a less active merger history, particularly at late times. Blizzard, on the other hand, has three more massive interactions \rev{with $\sim10^{10} \rm{M}_{\odot}$ halos at later times (the 50:1 and 40:1 interactions that precede the polar ring formation, and then a 60:1 merger).} Given that its second polar ring is destroyed, this also ends up populating its inner CGM with a significant amount of cool gas. This may be the reason why, despite forming an early thin disk, Blizzard starts significantly flaring at later times.

The other key difference between these two systems is the behavior of the relative orientation between the inner and outer disks. In Tempest, the orientation of the inner disk is relatively stable. The orientation of the outer disk, on the other hand, precesses significantly. In Blizzard, the opposite is true, although both components precess more significantly. This is likely a further consequence of the more populated local environment of Blizzard, as massive subhalo interactions can apply significant torques to the inner disk and dark matter halo.

\subsection{The Angular Momentum of Tempest's Polar Ring}\label{sec:angular_momentum}

Tempest's polar ring, while misaligned, is still a coherently rotating component of the system. Therefore, for it to precess towards the disk plane, a significant amount of torque is required. At $t=8.5$ Gyr ($z\sim0.5$), the polar ring has an average rotational velocity of 175 km s$^{-1}$ and a radius of $\sim$15 kpc. For this ring to rotate the 20\textdegree\ towards the disk plane between $t=7$ Gyr and $t=10$ Gyr, an average specific torque of 600 kpc km s$^{-1}$ Gyr$^{-1}$ is required. The orbiting subhalo seen in Fig.~\ref{fig:tempest_time_plots} has a peak mass of $6\times10^{9}$ \msun\ and distances ranging from $\sim10-100$ kpc, meaning that depending on orientation, it can exert a specific torque of up to 400-4000 kpc km s$^{-1}$ Gyr$^{-1}$.  \rev{At this time, the gravitational torque from an exponential disk fit to the gaseous surface density on a polar ring at an inclination of $\sim40$\textdegree, would be on the order of $\sim$100 kpc km s$^{-1}$ Gyr$^{-1}$.} The overall torque on the polar ring in this case is likely a combination of gravitational torques from this subhalo and the combined gravitational torques from the aligned disk plane and dark matter halo. \rev{As mentioned previously, the stellar profiles in FOGGIE are overly concentrated, and therefore will not significantly torque this polar ring. For a similar stellar mass, a stellar disk with a scale radius of 1-5 kpc would produce a torque between $\sim$100-450 kpc km s$^{-1}$ Gyr$^{-1}$, which would result in a more rapid precession of the polar ring.}

Once the polar ring begins transitioning to a warp and aligning with the inner disk, larger torques are required (Fig.~\ref{fig:tempest_ring_evolution}). Given that the subhalo has lost a significant amount of mass following its initial fly-by, this must be due to secular interactions with the disk, with hydrodynamical torques likely playing a significant role.

\subsection{Observations of warps and polar rings}

Most stellar warps have inclinations of a few degrees, with some reaching up to $\sim$10\textdegree\ \citep{reshetnikov25-warps}. Stronger warps have been detected in \HI\, up to $\sim 18$\textdegree\ \cite[e.g.,][]{Healy_2024-WarpedMhongooseDisk}, which is consistent with values we see in Tempest and Blizzard. Our sample only shows S-/integral shaped warps.
This may be a selection effect, as U-shaped warps may preferentially occur in group/cluster environments and are not often found in zoom-in simulations \citep{zee22-U_warps}.
A deeper analysis of the kinematic signatures of these warps, particularly in comparison with \citet{Healy_2024-WarpedMhongooseDisk} and \citet{lin2025-FEASTS-M51}, is presented in \citetalias{trapp25b}.

Galaxies with polar structures at a variety of angles ($\sim50-90$\textdegree) from the major axis of the disk have been observed \citep{Finkelman12-PolarRingObs}, although modeling the exact 3-D structure is difficult. These observations would be consistent with the variety of inclinations we see during the evolution of the polar ring structures in both Blizzard and Tempest. In the case of Tempest, the polar ring begins to incline to lower angles from the disk plane, although it begins transitioning into a warp at $\sim40$\textdegree. Polar rings are often observed around S0-like galaxies, although recent \HI\ observations have found candidates around other spiral systems \citep{deg23-HIPolarRings}.

There is still debate on the origin of both warps and polar rings. In this study, we see several different formation mechanisms for each. In Tempest, a merger with a satellite causes a polar ring to form, which then precesses into a warp. We also see a smaller warp form in the outer regions of the inner disk. Whether this is due to misalignments with the dark halo or tidal interactions is not clear. Importantly, the larger warp/polar ring from the merger event precesses to align itself with this warp in the inner disk over the course of $\sim 3$ Gyr. In Blizzard, we see the polar rings form primarily through tidal accretion of infalling satellites. The subsequent warp forms independently following the destruction of the polar ring from accretion of the leftover material in the CGM.

\subsection{Numerical Considerations}\label{sec:disc_numerical_considerations}

 As has been discussed in previous FOGGIE papers, these simulations employ exclusively thermal stellar feedback and do not account for sources of feedback other than Type II SNe. Similarly, these simulations do not model the effects of cosmic rays. These feedback sources can have significant effects on the structure of gas flows within the CGM \cite[e.g.,][]{hopkins21-CrDrivenOutflows} and disk \cite[e.g.,][]{trapp24}. It is unclear how significant stellar feedback may be in disrupting the structure of these extended, misaligned disks. Combined with the star formation scheme used in these calculations, this additionally results in overly-concentrated stellar profiles \citep{wright24}. As discussed in Sec.~\ref{sec:angular_momentum}, a stellar potential more distributed across the disk plane may more effectively torque misaligned components, potentially aligning them faster than what is seen in this study.

The FOGGIE simulations also do not include feedback from AGN, which may affect the growth of the galaxies at high redshifts \cite[e.g.,][]{somerville08-AgnFeedback,harrison17-AgnFeedback}. Similarly, these simulations do not have magnetic fields. While the global magnetic energy may be subdominant ($\beta>>1$), the magnetic field can be important locally due to local magnetic tension and thus can have a variety of consequences \citep[see, e.g.,][]{Ji_2018MNRAS.476..852J,Wibking_2025arXiv250610277W}.

We have found that local environmental effects -- the CGM and satellites -- have an influence on disk morphology. It is important to note that the galaxies in the FOGGIE simulations were chosen to have specific $z=0$ mass and relatively quiescent merger histories at late times ($z < 2$). This selection effect will have consequences for the local environment and for larger-scale neighborhood past $R_{\rm vir}$. This can locally impact the CGM density/pressure and, therefore, disk behavior in ways we cannot easily quantify with this selection, and the relatively small number of halos studied. 

Finally, the FOGGIE simulations focus resolution elements on the CGM at the cost of resolution within the disk. As discussed more in Appendix~\ref{sec:resolution_criteria}, we see that while the inner disks are not well-resolved, the median cell mass in the extended disk is comparable to other state-of-the-art cosmological simulations ($\sim10^{4}$~M$_{\odot}$). Even a few cells outside of our disk definition, this median cell mass drops several orders of magnitude, implying that the FOGGIE simulations are resolving the disk-halo interface better than other cosmological simulations. 

\section{Summary}

We investigate the evolution and observable properties of the extended \HI\ disks in Milky Way-mass galaxies using zoom-in simulations from the Figuring Out Gas \& Galaxies in Enzo (FOGGIE) simulation suite. We place the six halos on a continuum based on the amount of dense material in their CGM at late times: from \LessActive\ and \MoreActive. We generally find that these local environmental effects have a significant impact on the disk morphology and evolution. Our conclusions are as follows:

\begin{itemize}
    \item The mean and median column densities are similar within the disk; however, the median value drops off sharply at the disk edge, while the mean value is dominated by clumpy material. The \LessActive\ systems have steeper \HI\ profiles while the \MoreActive\ systems show a shallower profile at the disk edge (Fig.~\ref{fig:hi_curves}). 

    \item \LessActive\ systems tend to form thinner, coherently rotating disks and have hot inner CGMs near their virial temperature. \MoreActive\ systems show more chaotic evolution and tend to form thicker disks with distinctly misaligned components and have cooler inner CGM gas on average.
    
    \item The \MoreActive/\LessActive\ categorization is largely independent of virial mass, disk mass, or disk radius. Overall rotational support also evolves similarly (Fig.~\ref{fig:disorder_evolution}). Rather, we see correlations with CGM content (Fig.~\ref{fig:classification_quantification}) that imply influences from the local environment on disk scale height and the properties of the inner CGM.

    \item All extended disks in this study have misaligned features at some point during their evolution. The frequencies and lifetimes of these features vary significantly with the external influences of accreting CGM and satellites. 


    

    \item Two of our halos (Tempest and Blizzard) show similar warped morphology at $z=0$, but the formation and duration of their misaligned features vary due to environmental effects. Tempest has a long-lived polar ring that precesses towards the disk plane at a rate of $\sim 7$\textdegree\ per Gyr and transitions into a large warp. Blizzard forms two distinct polar rings from tidal accretion of interacting subhalos, but they are both destroyed after $\sim 1$\,Gyr. Following this, a warp begins to form. Despite these differences, the warps in Tempest and Blizzard are strikingly similar at $z=0$.





    \item{Resolution within the FOGGIE simulations is focused on resolving the CGM, however, the median mass resolution within the extended disks of these systems is comparable to other state-of-the-art cosmological hydro simulations ($\sim 10^{4} \, \rm M_{\odot}$). This median cell mass drops off sharply just a few kpc from our disk definition, implying the FOGGIE simulations are resolving the disk halo interface more robustly than other cosmological simulations (See Appendix~\ref{sec:resolution_criteria}).}

\end{itemize}

In this study, we find misaligned features are ubiquitous in our sample of Milky Way mass galaxies, with each system showing polar rings or warps at some point during its evolution. The origin, strength, and duration of these features vary strongly between systems, and are highly dependent on the local environment the central galaxy resides within. Interestingly, similar galactic morphologies can arise at $z=0$ from drastically different origins. \rev{We ultimately find a relation between the abundance of \HI\ within the CGM and the amount of substructure, which additionally relates to the presence of warps, polar rings, and disk settling}. This relation is uncorrelated with the total mass of the halo or disk. This implies that the local environment in which galaxies reside is significant for controlling how their \rev{central disks} develop and evolve, as opposed to being fully determined by the thermalization of the CGM.

\begin{acknowledgments}
We thank Lauren Corlies for pointing out that, with the density threshold set appropriately, the disk will just be a clump finder's largest ``clump''. 
We also thank the anonymous referee for their detailed and insightful comments that have helped improve the manuscript.

CWT was supported for this work in part by NASA via a
Theoretical and Computational Astrophysics Networks
grant \#80NSSC21K1053 and JWST AR \#5486.
VS was supported for this work in part by NASA via an
Astrophysics Theory Program grant \#80NSSC24K0772, HST AR \#17549, and HST GO \#17093.
CWT and VS were additionally supported by HST AR \#16151.
BWO acknowledges support from NSF grants \#1908109 and \#2106575, NASA ATP grants 80NSSC18K1105 and 80NSSC24K0772, and NASA TCAN grant 80NSSC21K1053.
Support for CL was provided by NASA through the NASA Hubble Fellowship grant \#HST-HF2-51538.001-A awarded by the Space Telescope Science Institute, which is operated by the Association of Universities for Research in Astronomy, Inc., for NASA, under contract NAS5-26555.
AA acknowledges support from the INAF Large Grant 2022 “Extragalactic Surveys with JWST” (PI Pentericci) and from the European Union – NextGenerationEU RFF M4C2 1.1 PRIN 2022 project 2022ZSL4BL INSIGHT.
RA acknowledges funding from the European Research Council (ERC) under the European Union's Horizon 2020 research and innovation programme (grant agreement 101020943, SPECMAP-CGM).

Computations described in this work were performed using the publicly available \textsc{Enzo} code (\href{http://enzo-project.org}{http://enzo-project.org}), which is the product of a collaborative effort of many independent scientists from numerous institutions around the world. Their commitment to open science has helped make this work possible. The Python packages {\sc matplotlib} \citep{hunter2007}, {\sc numpy} \citep{walt2011numpy}, {\sc ROCKSTAR} \citep{Behroozi2013a}, {\sc tangos} \citep{pontzen2018}, \textsc{scipy} \citep{scipy2020}, {\sc yt} \citep{ytpaper}, and {\sc Astropy} \citep{astropy2013,astropy2018,astropy2022} were all used in parts of this analysis or in products used by this paper. 

Resources supporting this work were provided by the NASA High-End Computing (HEC) Program through the NASA Advanced Supercomputing (NAS) Division at Ames Research Center and were sponsored by NASA's Science Mission Directorate; we are grateful for the superb user-support provided by NAS. Resources were also provided by the Blue Waters sustained-petascale computing project, which is supported by the NSF (award numbers ACI-1238993 and ACI-1514580) and the state of Illinois. Blue Waters is a joint effort of the University of Illinois at Urbana-Champaign and its NCSA. Computations described in this work were performed using the publicly-available Enzo code, which is the product of a collaborative effort of many independent scientists from numerous institutions around the world. 

\end{acknowledgments}

\begin{contribution}

CWT led the analysis and writing of the paper. MSP, BWO, and JT  are the principal investigators of the FOGGIE collaboration; they obtained funding for, developed, and ran the simulations used in this work, and provided valuable feedback and insights throughout the project. CL characterized halo masses and star formation rates, as well as contributing to the interpretation of the results. ACW characterized satellite evolution and provided feedback on the draft. BDS contributed through discussions and feedback on the draft and developed the initial conditions for the FOGGIE simulations. VS assisted in figure design and provided feedback on the draft. RCS, AA, and RA contributed through discussions during the analysis phase and feedback on the draft.


\end{contribution}

%



\appendix

\section{Effects of Simulated Resolution} \label{sec:simulated_resolution}

In this section, we investigate how the spatial resolution of the simulation affects the structure of the \HI\ disks. We focus on two main topics. The first topic is how varying simulated resolution affects the size of the \HI\ disk, the sharpness of the disk edge, and the amount of detectable \HI\ outside the main galaxy. The second topic is how the gas in these systems is being refined, whether by cooling refinement, density refinement, or forced spatial refinement.

\subsection{Changes to the \HI\ profile at different resolutions} \label{sec:covering_fractions}
\begin{figure}[ht!]
\begin{centering}
\includegraphics[width=0.475\textwidth]{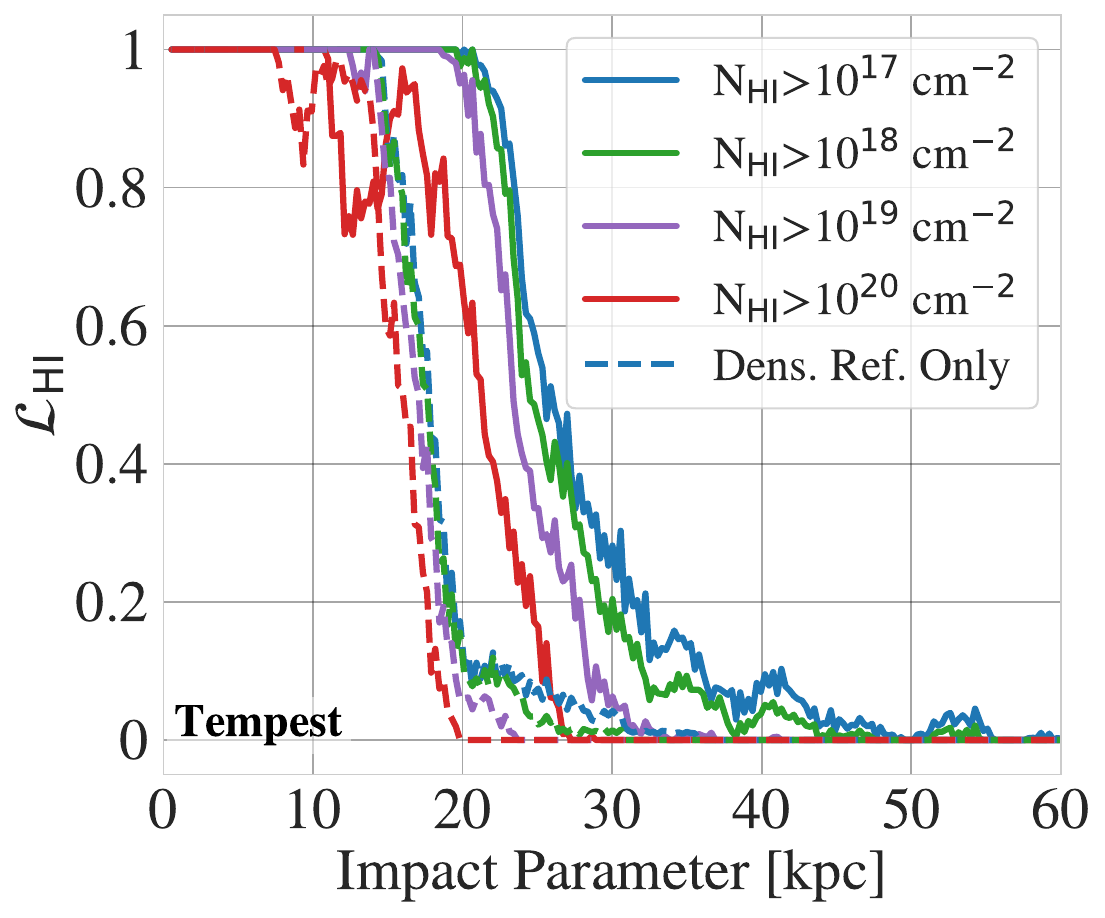}
\caption{
The \HI\ covering fraction for Tempest as a function of impact parameter in face-on projections. The solid lines show the covering fractions at various column density cutoffs for the fiducial runs. The dashed lines show the same covering fraction profiles for the same runs without cooling or forced refinement. In the fiducial runs, the disks are larger by about 10 kpc, show a more gradual drop off as a function of radius, and show significant clumpy material outside the main disk.
\label{fig:tempest_covering_fraction}}
\end{centering}
\end{figure}

Fig.~\ref{fig:tempest_covering_fraction} shows the face-on \HI\ covering fraction (\LHI) for Tempest at four different column density thresholds. The solid lines show the fiducial FOGGIE refinement scheme (cooling, density, and forced refinement), while the dashed lines show only density refinement. There are three main results.

Firstly, in the runs with no forced refinement and no cooling refinement, once the \HI\ covering fraction drops outside the main disk, it stays at approximately zero for all column density thresholds. This is simply because no clumpy structure is present in this run, as it is not resolved.

Secondly, the disks run with the additional refinement are larger by around 10 kpc. This trend has been seen in some simulation suites that vary resolution \cite[e.g.,][]{hopkins18-FIRE2}, but some simulations show large disks even with relatively low resolution \cite[e.g.,][]{diemer19}. This implies that, while this may be a general effect of resolution on an individual system, the details of galaxy growth and disk formation are complex, and subject both to the environment and characteristics of the individual system, as well as the various numerical techniques and subgrid prescriptions that go into a given simulation suite.

Finally, the drop-off in \LHI\, is less sharp in the disks run with additional refinement. This is not unexpected, as increasing CGM resolution has been shown to allow smaller-scale, higher-density substructure to be resolved \citep{vandevoort19, hummels19}. This change in slope is particularly significant, as it implies that the disk-halo interface and the sharpness of the disk edge may be highly sensitive to simulation resolution. Similarly, the structure and column density of this material have important consequences for the observability of this interface, as discussed more in \citetalias{trapp25b}.

\subsection{Where gas is under/over resolved} \label{sec:resolution_criteria}

\begin{figure}[ht!]
\includegraphics[width=0.49\textwidth]{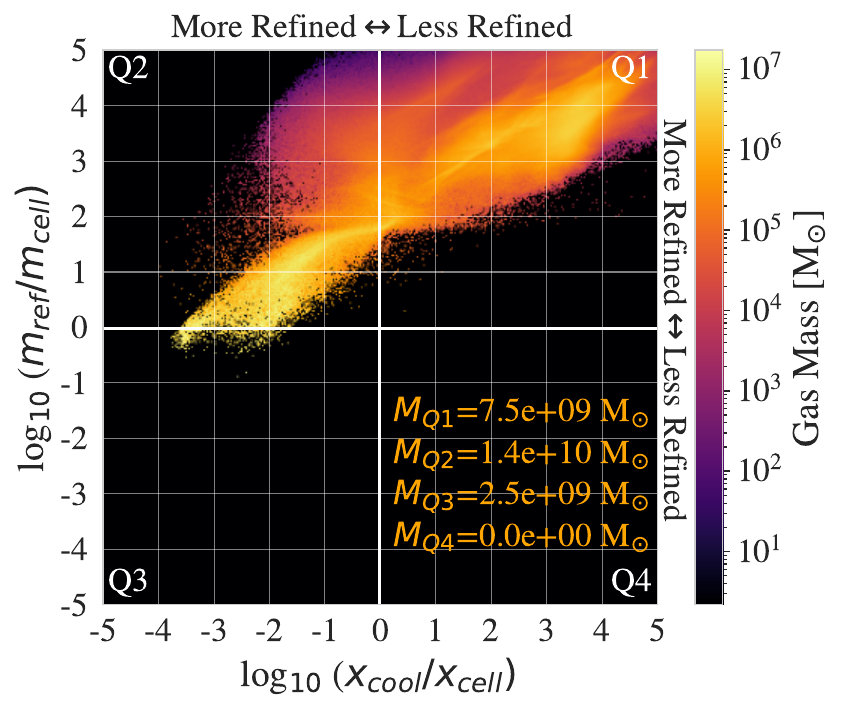}
\caption{
Refinement criteria for each cell as calculated by its cooling length ($x_{cool}$) and mass ($m_{ref}$) compared to the actual refinement length ($x_{cell}$) and mass of the cell ($m_{cell}$). The 2-D histogram shows the relative refinement of cells in Tempest at $z=0$, color-coded by mass. Only cells in the forced refinement region are shown. In the bottom right, the total gas mass in each quadrant is shown. There are no cells in the bottom right quadrant (Q4), indicating that all cells in the forced refinement regions that are resolved in terms of cooling length are also resolved in terms of mass.
\label{fig:tempest_refinement_quadrants}}
\end{figure}

\begin{figure}[ht!]
\begin{centering}
\includegraphics[width=0.45\textwidth]{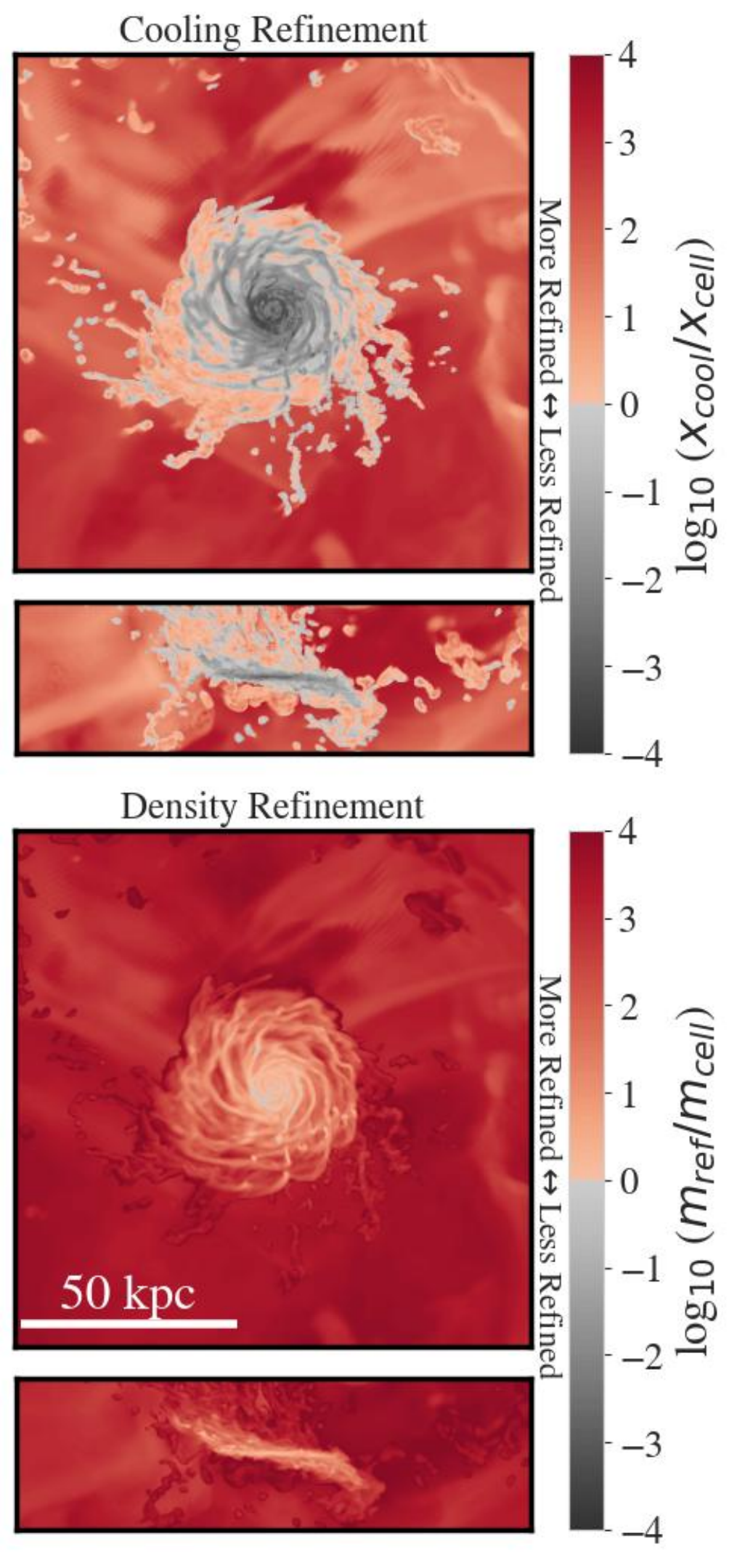}
\caption{
Face on projections for gas in the refinement box of Tempest at $z=0$, color-coded by the refinement criteria introduced in Fig.~\ref{fig:tempest_refinement_quadrants}. Projections are weighted by \HI\ density. \textit{Top:} Projections of the cooling refinement criteria. Positive (red) values are more refined than this criterion, while negative (gray) values are less refined. Note the edges of clumps and the disk edge have values near or below 0, implying rapid cooling.  \textit{Bottom:} Projections color-coded by the density refinement criterion. Most of the gas in the system is more refined than this criterion, except for the disk center.
\label{fig:tempest_refinement_projections}}
\end{centering}
\end{figure}

\begin{figure}[ht!]
\begin{centering}
\includegraphics[width=0.45\textwidth]{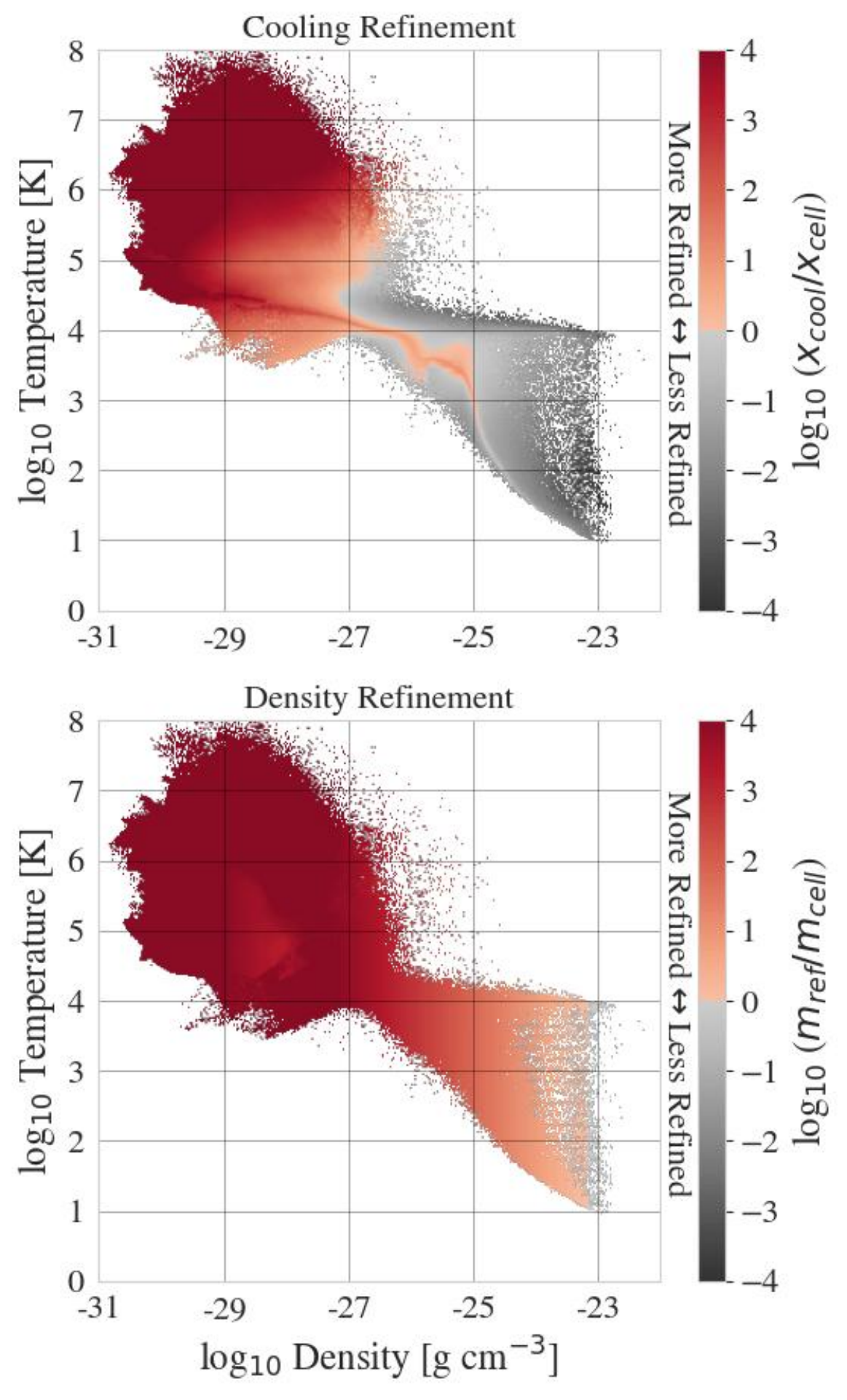}
\caption{
Phase diagrams for gas in the refinement box of Tempest, color-coded by the refinement criteria introduced in Fig.~\ref{fig:tempest_refinement_quadrants}. \textit{Top:} Phase diagrams color-coded by the cooling refinement criteria. Positive (red) values are more refined than this criterion, while negative (gray) values are less refined. \textit{Bottom:} Phase diagrams color-coded by the density refinement criteria. The density criterion is largely dependent on density and does not depend strongly on temperature.
\label{fig:tempest_refinement_phases}}
\end{centering}
\end{figure}

\begin{figure}[ht!]
\includegraphics[width=0.49\textwidth]{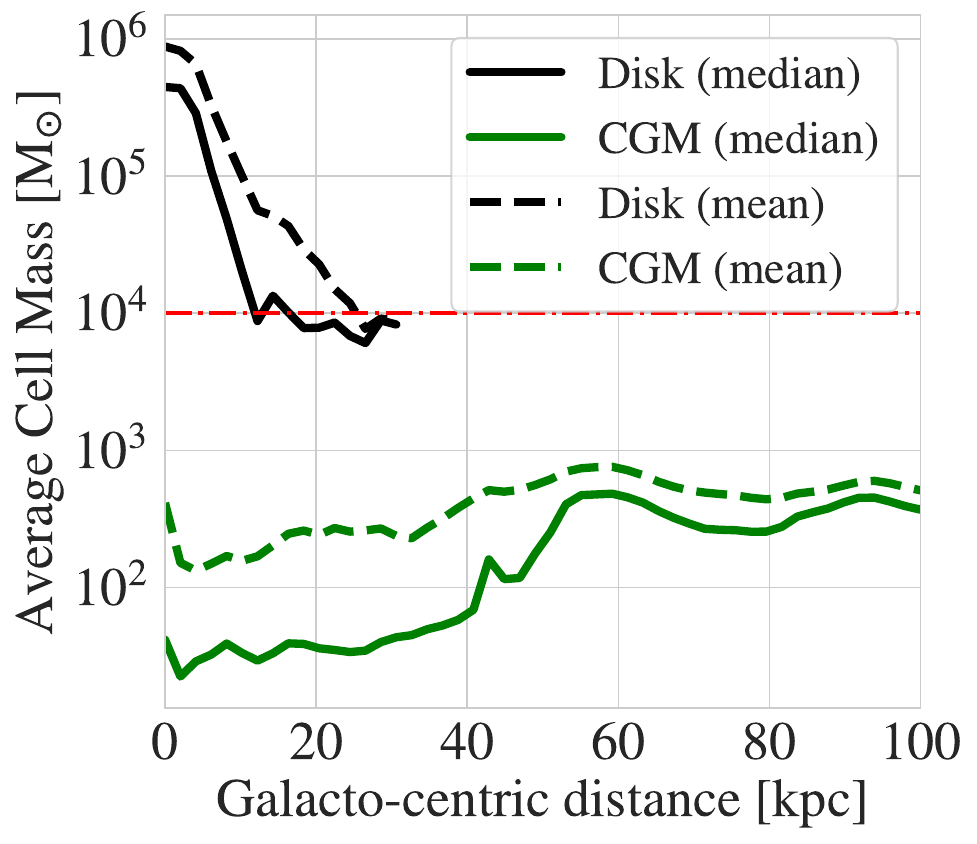}
\caption{
Average cell masses within our disk definition (black) and the CGM (green) for Tempest. The solid lines show median values and the dashed lines show mean values. Cell masses within the disk are large, but the median values in the extended disk are approaching $10^{4}$ \msun (red, dash-dotted line), a value typical of other cosmological hydrodynamical galaxy simulations that are not focused on resolving the CGM. Cell mass in the CGM is orders of magnitude smaller. Other halos show the same trend.
\label{fig:tempest_cell_masses}}
\end{figure}

\begin{figure}[ht!]
\includegraphics[width=0.49\textwidth]{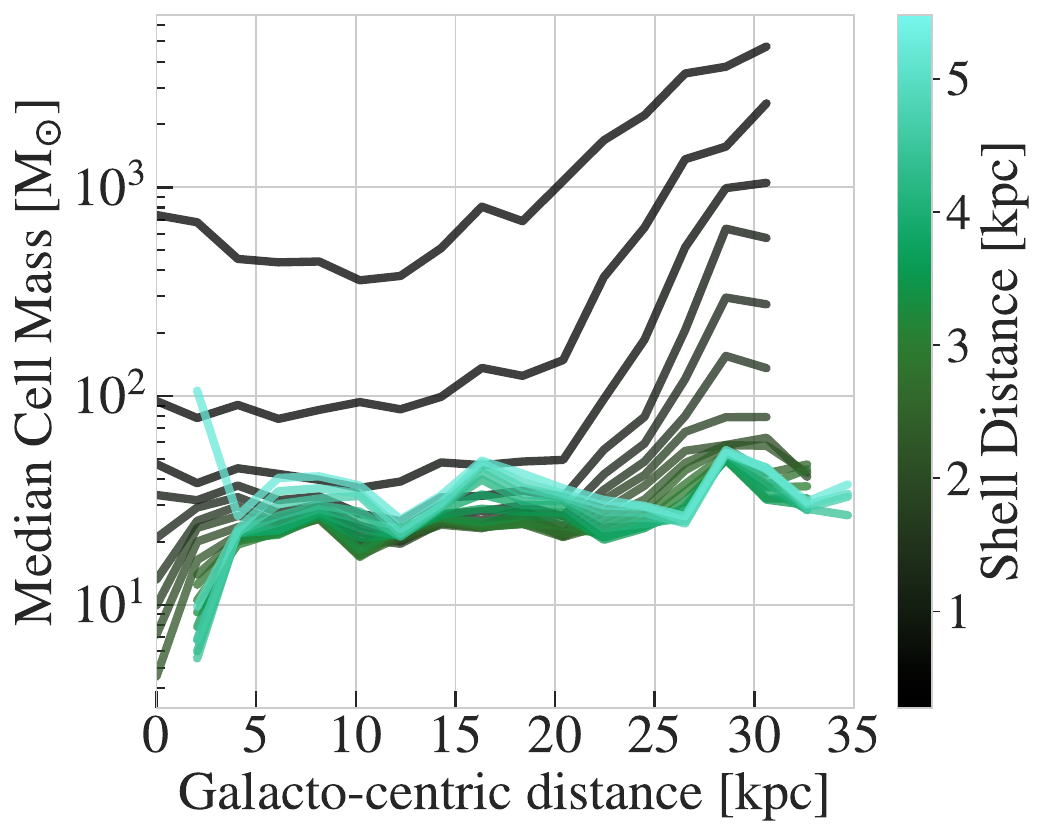}
\caption{
Median cell masses within 0.274 kpc (1 cell) thick shells around the disk definition. Shells were identified by a binary dilation from the disk surface. Each shell only includes cells outside the definition of the previous shell (and disk). Median cell mass drops below 100 \msun\ within a few shells.
\label{fig:tempest_shell_cell_masses}}
\end{figure}

The FOGGIE simulations use three schemes for applying high grid resolution where it is most needed: density refinement, cooling refinement, and forced refinement. The density refinement, which is common to all Adaptive Mesh Refinement (AMR) methods, subdivides a cell if its mass ($m_{cell}$) is above a certain threshold ($m_{ref}$), up to a maximum refinement level of 11. Similarly, the cooling refinement subdivides a cell if its size ($x_{cell}$) is greater than the cooling length of the gas ($x_{cool}$), also up to a maximum refinement level of 11. Finally, the forced refinement scheme imposes a minimum refinement level of 9 within a 288 kpc comoving box, ensuring there is a maximum cell size of 1100 pc (comoving) within the CGM (see Sec.~\ref{sec:simulations}). The cooling refinement criterion is imposed only within this moving box centered on the galaxy host. 

Fig.~\ref{fig:tempest_refinement_quadrants} shows a 2-D histogram of how many cells are more refined or less refined than the ideal mass/density refinement scheme and the cooling refinement scheme. Only cells within the moving forced refinement box are considered in this plot. For example, a cell with a ratio of $m_{ref}/m_{cell} > 1$ is more refined than it needs to be based only on the mass/density criterion. This will occur when the cell is pushed to a higher refinement level by one of the other two criteria. The only cases in which a cell is less-refined than a given refinement criterion are when it is at the maximum refinement level, and thus is not allowed to refine any more.

Gas in the top right (Q1) is more-refined by both criteria. This means the cell size is smaller than both the cooling length and the length scale determined by the cell's density.  Gas in the top left (Q2) is less-refined in cooling, but more-refined in mass. This corresponds to cells that are not overmassive, but have cooling lengths smaller than the cell size. Gas in the bottom left (Q3) is less-refined by both criteria, meaning it is over-massive and has cooling lengths smaller than the cell size. Gas in the bottom right (Q4) is more-refined in cooling, but less-refined in mass. There is no gas within the forced refinement region that falls in this quadrant; however, there are cells outside this region that do. This implies that within the forced refinement region, the cooling refinement criterion is triggered more often than the density refinement. The mass of each quadrant is displayed in the bottom right of the plot. Most of the gas mass is in Q2 (less-refined by cooling, more-refined by mass). Despite the relatively small number of bins in this plot, there is still a substantial amount of mass in Q3 (less-refined in both).

Fig.~\ref{fig:tempest_refinement_projections} shows the density-weighted projections of these two refinement criteria (density and cooling length). The plot on the top shows the cooling refinement projection. The gas less refined than the cooling refinement is largely present in the denser, spiral-like structure within the disk itself. It also traces the edges of the extended disk and the edges of clumps. The plot on the bottom shows the density refinement projection. Most gas is more refined than this criterion, by design, except for gas at the dense center of the disk. This implies there is rapid cooling at the edges of the disk and the edges of the clumps, as would be expected \citep{fielding17_cooling,ji19-cooling,fielding20_cooling}. Additional resolution could be focused on these regions. Resolving the cooling lengths within the disk would require many more levels of refinement and is not currently computationally feasible, while our runs are also simultaneously resolving the CGM.

\begin{figure*}[ht!]
\includegraphics[width=.95\textwidth]{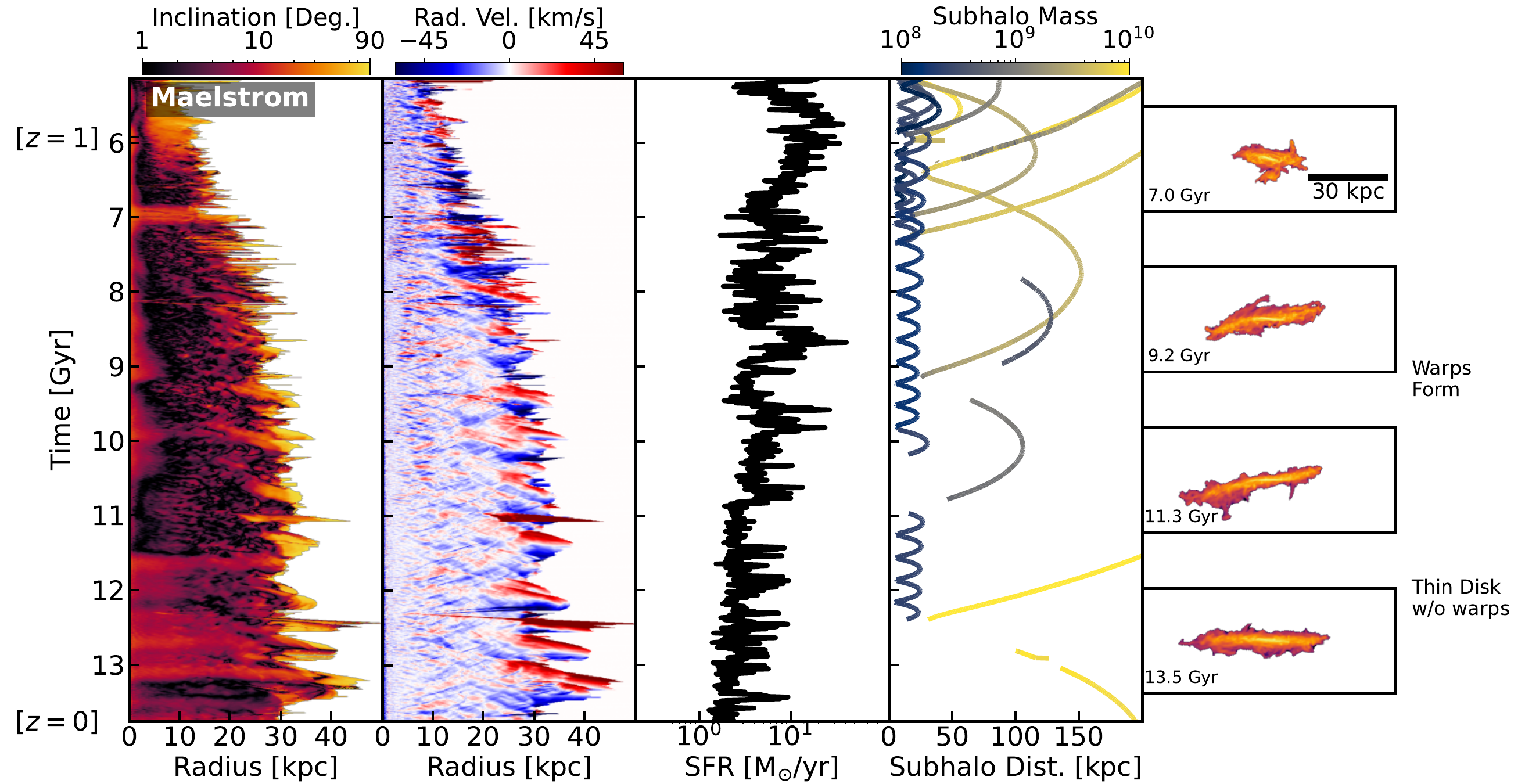}
\caption{
Same as Fig.~\ref{fig:tempest_time_plots}, but for Maelstrom. This system features a somewhat active early merger history, but rapidly settles to form a thin disk. A few warps are briefly excited throughout its history, but do not last for longer than $\sim1$ Gyr typically. \label{fig:maelstrom_time_plots}}
\end{figure*}

Fig.~\ref{fig:tempest_refinement_phases} shows the phase diagrams for Tempest color-coded by the two refinement criteria. The top plot is color-coded by the cooling refinement criteria. There is a transition from more-refined to less-refined at $\rho\sim10^{-27} \rm{g~cm^{-3}}$ and $T\sim10^4-10^5 K$, around the peak of the cooling curve. A population of more-refined gas exists at higher temperatures and densities, corresponding to gas in the outer regions of the disk (light red regions in Fig.~\ref{fig:tempest_refinement_projections}). The lower plot is color-coded by the density refinement criterion. This criterion depends almost entirely on density. Gas starts to become less resolved than this criterion around $\rho=10^{-25}$ g cm$^{-3}$ and is fully under-resolved by $\rho=10^{-23}$ g cm$^{-3}$.

Fig.~\ref{fig:tempest_cell_masses} shows the median (and mean) cell masses within our disk definition (Sec.~\ref{sec:hi_disk_definition}) and within the CGM for Tempest. Of particular note is the fact that the median cell mass in the extended disk reaches $10^{4}~\rm{M}_{\odot}$. This is a comparable resolution to other cosmological hydrodynamical simulations that are not focused on resolving the CGM \cite[e.g.,][]{hopkins18-FIRE2,pillepich19-TNG50,applebaum21}. Similarly, Fig.~\ref{fig:tempest_shell_cell_masses} shows the median cell masses of 0.274 kpc (one cell) thick shells around our disk definition. There is significant evolution within the first $\sim 1$ kpc away from the disk, but even the first shell is well below $10^{4}$\msun. Median cell masses quickly reach values of less than $10^{2}$ \msun. This, combined with the previous figures showing we are resolving the outer half of the disk by both the mass- and the cooling-refinement criteria, implies the FOGGIE simulations are adequately resolving these extended disks and, importantly, resolving the disk halo interface more robustly than other cosmological simulations.

\section{Additional Timeseries}\label{sec:other_halo_time_evolutions}

\begin{figure*}[ht!]
\includegraphics[width=.95\textwidth]{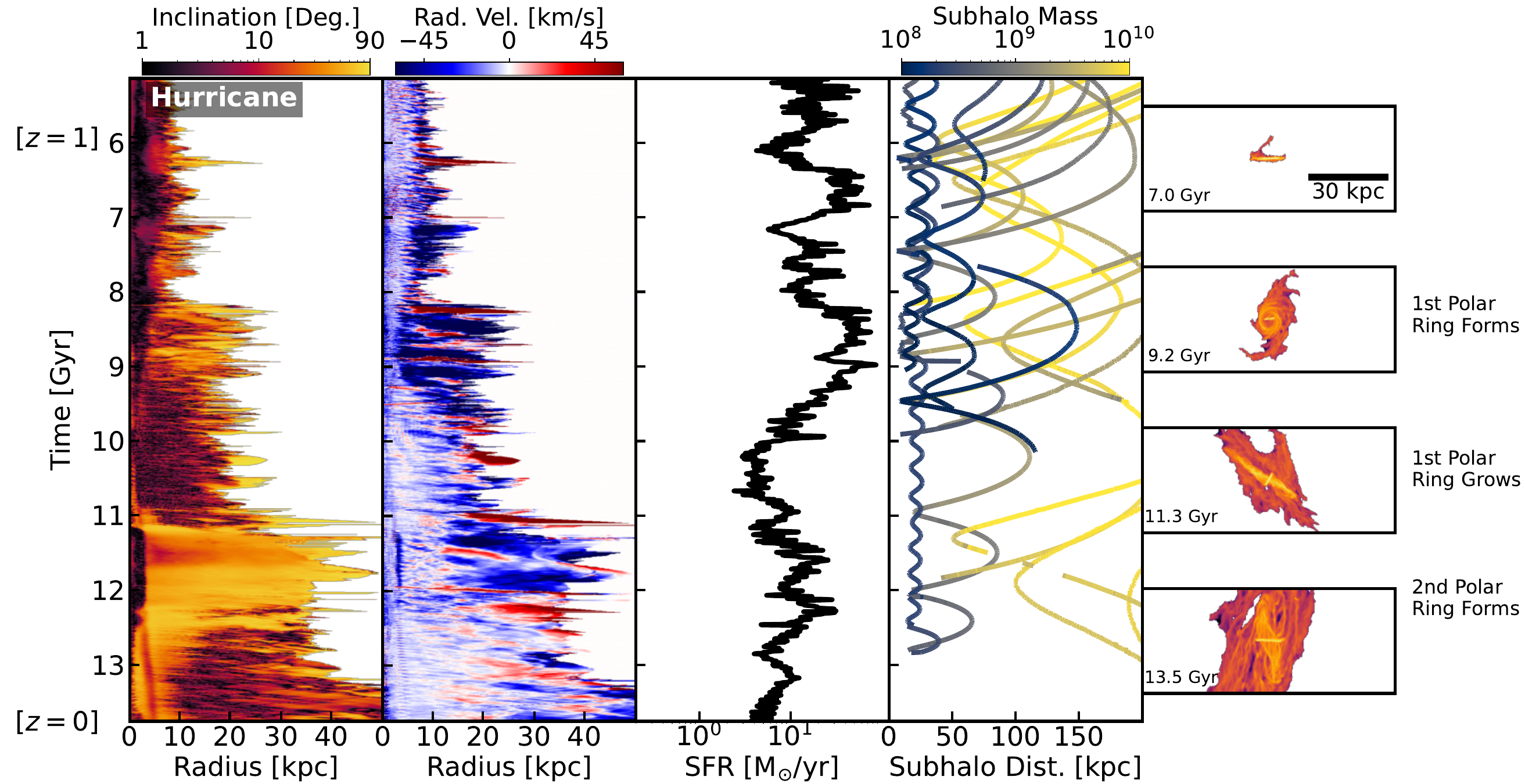}
\caption{
Same as Fig.~\ref{fig:tempest_time_plots}, but for Hurricane. Hurricane features a much more active merger/interaction history and therefore has elevated SFRs and three distinct components (two polar rings and a contracted inner disk). \label{fig:hurricane_time_plots}}
\end{figure*}

\begin{figure*}[ht!]
\includegraphics[width=.95\textwidth]{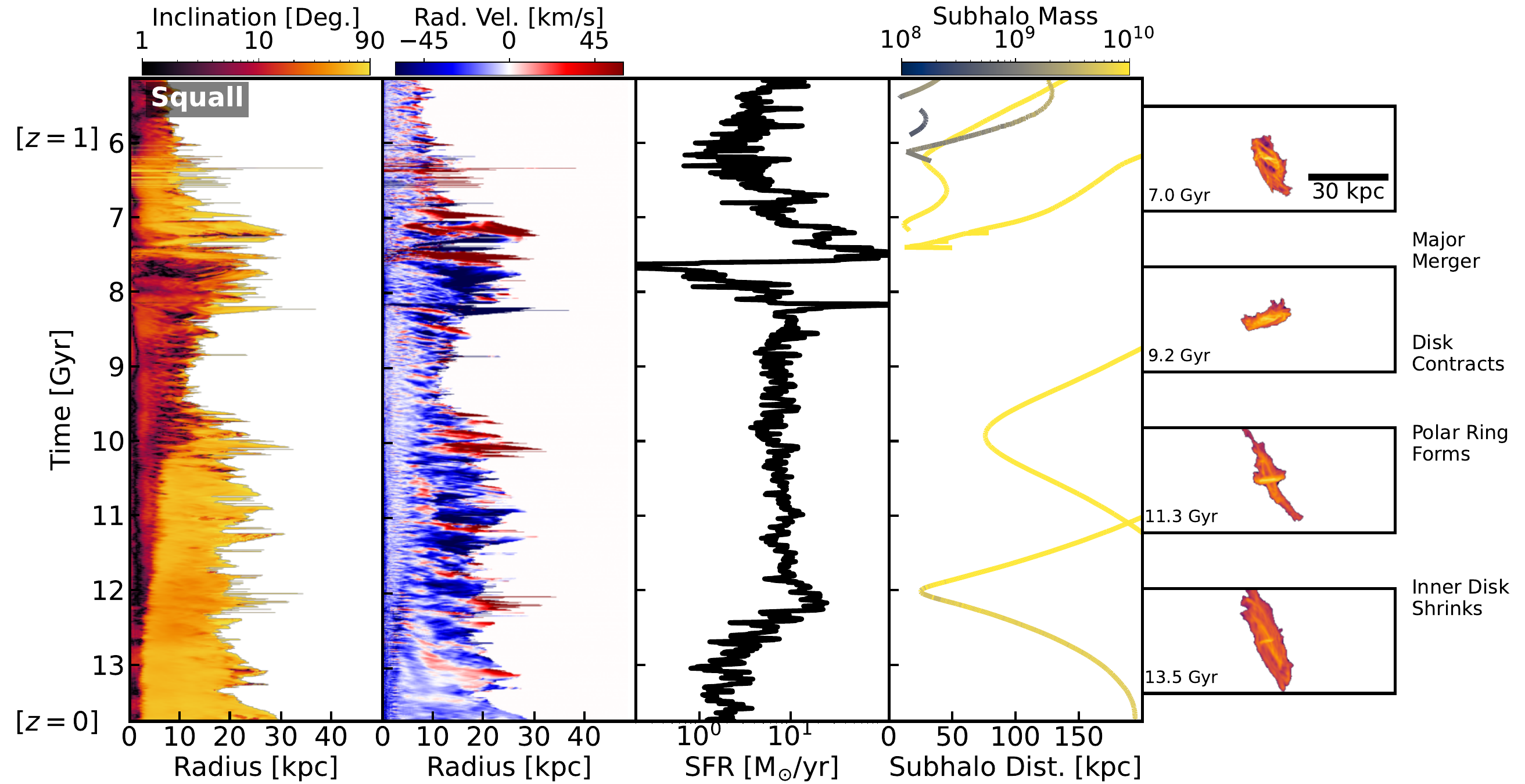}
\caption{
Same as Fig.~\ref{fig:tempest_time_plots}, but for Squall. This system has two relatively high mass mergers between $t\sim7$-8 Gyr, resulting in a relatively large disk that rapidly contracts. A polar ring forms around $t\sim9.5$ Gyr, which becomes the dominant component of the galaxy as the inner disk contracts. \label{fig:squall_time_plots}}
\end{figure*}

In the following subsections, we present versions of Fig.~\ref{fig:tempest_time_plots} for Maelstrom, Hurricane, and Squall. We briefly describe the history and morphology of each system, in the context of the distinction between the \LessActive~and \MoreActive~systems. All systems show misaligned features at some points during their evolution. Maelstrom shows only minor warps that quickly dissipate relative to the other systems in our sample. Both Squall and Hurricane show dramatic polar ring structures, with compact inner disks.

\subsection{The Rapidly Forming Thin Disk of Maelstrom}
Fig.~\ref{fig:maelstrom_time_plots} shows the evolution of Maelstrom between $z=1$ and $z=0$. Maelstrom is classified as a \LessActive~galaxy, and has a slightly more populated CGM than Tempest. Despite featuring numerous mergers and interactions around $z=0$, Maelstrom rapidly settles and forms a thin disk. Unlike other systems in this redshift regime, there is no obvious transition between `bursty' and `stable' star formation, with the star formation rate dropping relatively gradually from 10 \Msunperyear\ to a few \Msunperyear. While relatively small warps (compared with Tempest and Blizzard) can be seen throughout the evolution of Maelstrom, they typically do not last longer than $\sim1$ Gyr before dissipating. By $z=0$, there are no remaining warps. The radial velocity profile of Maelstrom is similar to the inner disk of the other systems, although the disk edges still show higher average radial speeds.

The lack of any polar ring/disk features during Maelstrom's evolution sets it apart from the other systems in our sample. This may be due to the fact that the mergers/interactions around $z=0$ are not as massive as some of the others. It may also be due to the specific details and geometry of how these mergers occur that prevent a coherent polar feature from forming.

\subsection{The Double Polar Rings of Hurricane} \label{sec:hurricane_time_evolution}
Fig.~\ref{fig:hurricane_time_plots} shows the evolution of Hurricane between $z=1$ and $z=0$. Hurricane is one of the \MoreActive~systems and features a much more active merger/interaction history than any other system. Correspondingly, it has elevated SFRs and more misaligned components, showing three distinct subcomponents at $z=0$.

At early times ($t<8$ Gyr), several misaligned components are formed and dissipate on relatively short time scales. At $t \sim 8$ Gyr ($z\sim0.6$), a polar ring is formed and the inner disk contracts. At $t \sim 11$ Gyr ($z\sim0.2$), this polar ring grows significantly. At $t \sim 12$ Gyr ($z\sim0.1$), a second polar ring forms. The system becomes more stable in terms of radial gas transport and star formation rate once the three components stabilize.

\subsection{The Shrinking Inner Disk of Squall}

Fig.~\ref{fig:squall_time_plots} shows the evolution of Squall between $z=1$ and $z=0$. Squall is one of the \MoreActive~systems, not forming a thin disk at $z=0$. This system has two relatively high mass mergers between $t\sim7$-8 Gyr ($z\sim0.8$ to $z\sim0.6$), one of which is a 2:1 merger. This results in a burst of star formation that creates a relatively large, diffuse disk. This leads to a brief period of quenching before the disk rapidly contracts, and the star formation rate stabilizes.

Around $t\sim9.5$ Gyr ($z\sim0.4$), a polar ring is formed, likely due to accretion from the interacting, massive subhalo. This polar ring becomes the dominant component of the galaxy, as the inner disk continues to shrink until it reaches a relatively stable size by $t\sim12$ Gyr ($z\sim0.1$). This contraction of the inner disk is correlated with high inward radial velocities and SFRs. 

Following the major merger at $t\sim7$ Gyr ($z\sim0.8$), there are a few other interactions with massive halos that likely populate the CGM with a large amount of cold, dense gas. Following the last interaction at $t=12$ Gyr ($z\sim0.1$), the gas mass of the central galaxy decreases (Fig.~\ref{fig:disk_mass_evolution}), as well as the star formation rate. This may eventually lead to a more coherent disk structure, but it is unclear, as there is still a large amount of material in the CGM.

\section{Exploring the \HI\ Disk Definition}\label{sec:disk_definition_appx}

In this section, we discuss in more detail the choices made in our disk identification algorithm (see Sec.~\ref{sec:hi_disk_definition}). We specifically present how the holes and ``donut'' holes are filled within the disk. We also discuss the choices for the initial \HI\ number density cutoff ($n_{cgm}$) and the unitless constant $\gamma$ presented in Eq.~\ref{eq:disk_cutoff}. 

\subsection{Hole Filling}

\begin{figure}[ht]
\begin{centering}
\includegraphics[width=.45\textwidth]{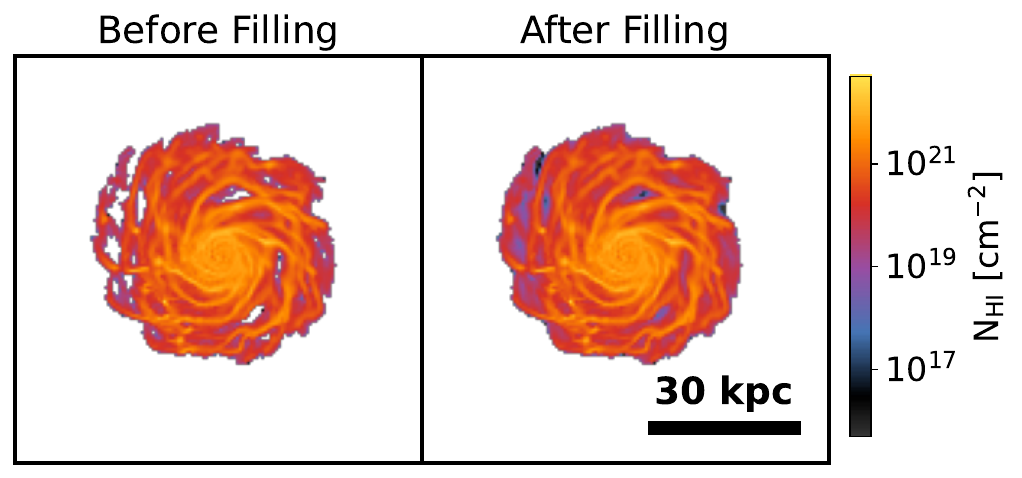}
\caption{\HI\ column density projections for the disk definition of Tempest at $z=0$ before (\textit{left}) and after (\textit{right}) the hole filling steps described in Sec.~\ref{sec:hi_disk_definition}. In brief, this process fills any true topological holes (fully enclosed) through a binary fill operation, and any ``donut'' holes that pierce the disk through a binary closing operation.
\label{fig:hole_filling}}
\end{centering}
\end{figure}

Fig.~\ref{fig:hole_filling} shows \HI\ column density projections of our disk definition before and after the two hole-filling steps. In brief, the first step fills any true topological holes (fully enclosed) through a simple binary fill operation. The second step fills any ``donut'' holes, or holes that fully pierce the disk through a binary closing operation \citep{petros87-BinaryClosing}. This operation takes the disk mask as a binary array and first applies a binary dilation operation, increasing the size of the mask at the edges. This is followed by a binary erosion operation, which decreases the size of the mask at the edges. If a donut hole is filled by this first step, it will not be eroded by the second step, resulting in a binary mask with filled holes. The kernel size (7 kpc) determines how large the dilation/erosion operations are, and thus, the size of the holes that are filled.

As can be seen in Fig.~\ref{fig:hole_filling}, this effectively fills in holes within the disk with low column density material. While this operation has minimal effects on the rest of the disk mask, the edge of the disk mask is filled in slightly. This is the result of ``near'' holes, or empty regions that are nearly encircled (i.e., surrounded on 3 sides instead of 4) being filled. This effect becomes more significant for larger kernel sizes, so an arbitrarily large size should not be selected.

\subsection{Disk Parameterization Choices}
\begin{figure}[ht!]
\begin{centering}
\includegraphics[width=.45\textwidth]{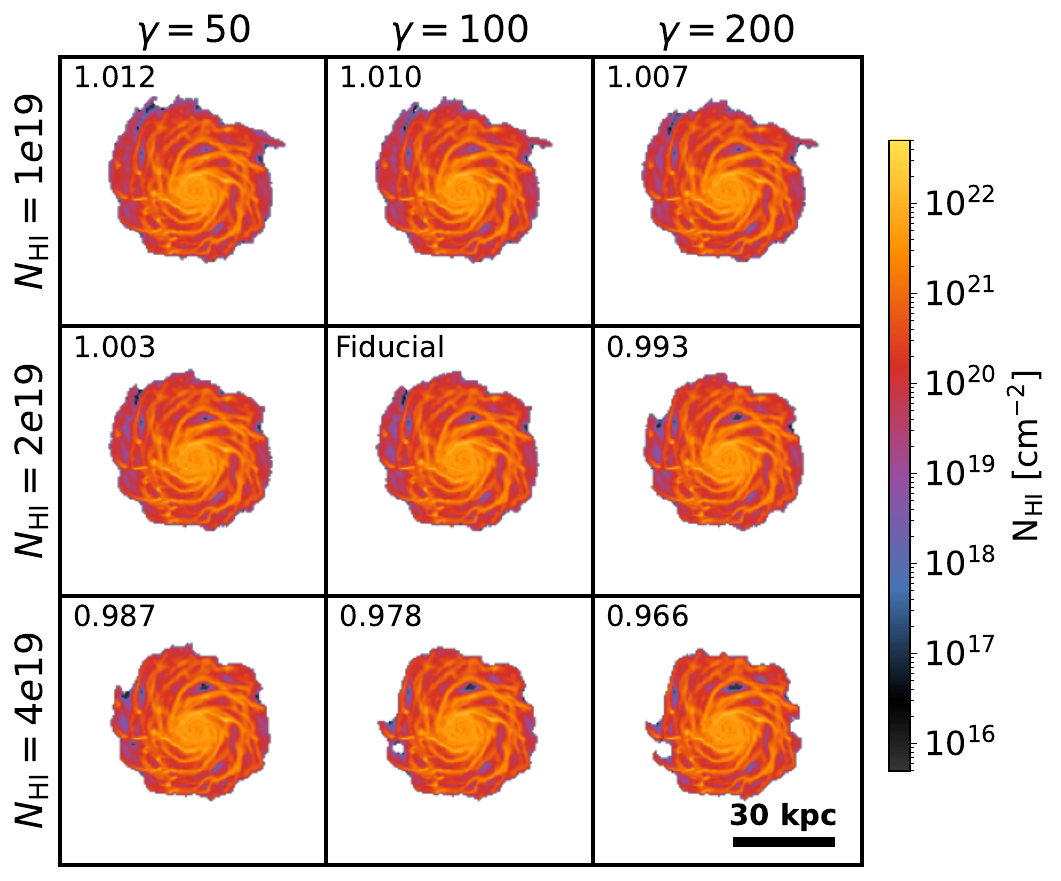}
\caption{\HI\ column density projections for disk definition for Tempest at $z=0$ for various parameterizations of Eq.~\ref{eq:disk_cutoff}. The \HI\ column density cutoffs ($N_{\rm HI}$) are used to set $n_{cgm}$, while $\gamma$ multiplies the standard deviation of the CGM \HI\ number density to increase $n_{\rm cut}$ in dense environments. The value in the top left corner of each image corresponds to the ratio of \HI\ mass to our fiducial disk definition (\textit{center}, $\gamma=100$, $N_{\rm HI}=2\times10^{19}~\rm{cm}^{-2}$). There is little change to the overall morphology and \HI\ mass in the disk. \label{fig:disk_parameters_z0}}
\end{centering}
\end{figure}

\begin{figure}[ht!]
\begin{centering}
\includegraphics[width=.45\textwidth]{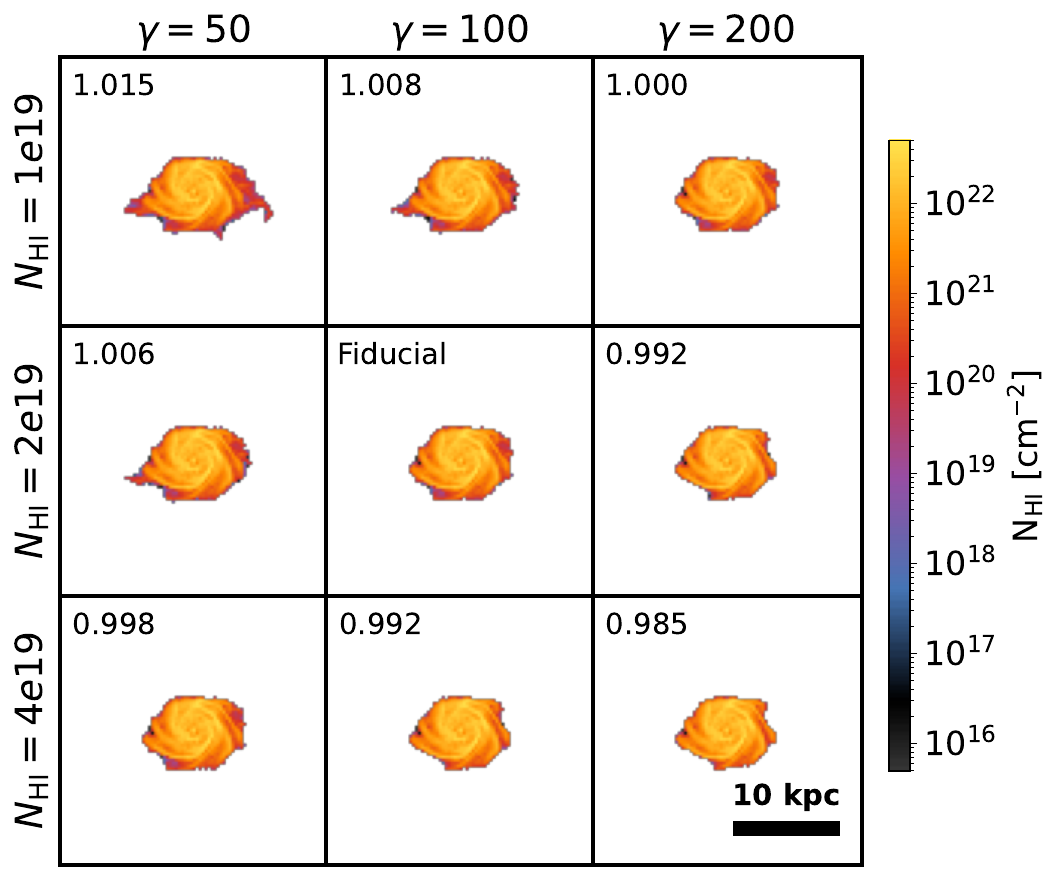}
\caption{Same as Fig.~\ref{fig:disk_parameters_z0}, but for Tempest at $z=1$ disk definition for Tempest at $z=1$. Although $\gamma$ can be more significant for the denser environments at higher redshift, there is still little variation in disk morphology and \HI\ mass. \label{fig:disk_parameters_z1}}
\end{centering}
\end{figure}

As a reminder, Eq.~\ref{eq:disk_cutoff} shows how we set the \HI\ number density threshold used to identify the disk as follows:
\begin{equation} 
    n_{\rm cut} = \frac{n_{\rm cgm}   + \gamma\sigma_{\rm cgm}}{a^{3}}~,
\end{equation}
where $\sigma_{\rm cgm}$ is the volume-weighted standard deviation of the \HI\ number density of the CGM gas as identified by the initial density cut ($n_{\rm cgm}$) and within the virial radius ($R_{\rm vir}$), $a$ is the scale factor of the universe, and $\gamma$ is a unitless constant. We set $n_{\rm cgm}$ such that a 1 kpc thick column of gas would have some given \HI\ column density $N_{\rm HI}$. Our fiducial choices for the two free parameters are: $N_{\rm HI}=2\times10^{19}~\rm{cm}^{-2}$ and $\gamma=100$.

Fig.~\ref{fig:disk_parameters_z0} and Fig.~\ref{fig:disk_parameters_z1} show the disk definition of Tempest for various values of these two parameters at redshifts $z=0$ and $z=1$, respectively. Tempest was chosen as it is the least populated system at $z=0$, while still residing in a dense local environment at $z=1$. This makes it a good example of the two extremes this definition must satisfy.

Ultimately, as shown in Fig.~\ref{fig:hi_curves} and discussed in Sec.~\ref{sec:hi_disk_definition}, \HI\ density falls off sharply at the disk edge, so our disk definition is not very sensitive to the specific parametrization in this range. Even for the \MoreActive\ systems, the edge of the disk still tends to be sharply delineated, even if there is more non-contiguous clumpy material in the inner CGM. This is illustrated by both the minimal changes to disk morphology and the fact that the total \HI\ mass in the disk definition varies only by a few percent or less, shown in the top left of each panel in Fig.~\ref{fig:disk_parameters_z0} and Fig.~\ref{fig:disk_parameters_z1}.

\subsection{Time Evolution of Disk Definition} \label{sec:disk_size_evolution}

\begin{figure}[ht!]
\includegraphics[width=0.48\textwidth]{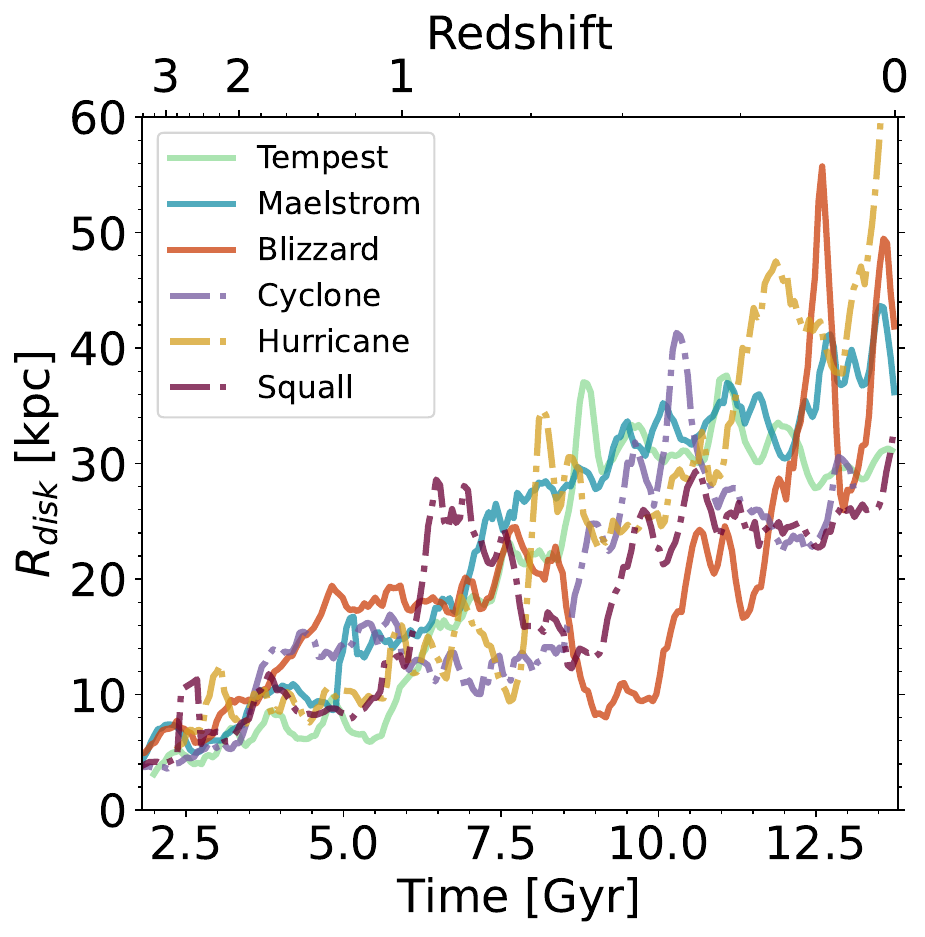}
\caption{
Evolution of the maximum radial extent of each disk, based on our disk definition. All systems grow similarly over cosmological time, despite classification (denoted by line type). Tempest and Maelstrom grow more or less monotonically, while the other systems go through periods of rapid growth and contractions. There is no obvious signal in disk growth that could be associated with the \LessActive\ or \MoreActive\ categories. Disk radius in this case is defined by the radius of the gas cell furthest from the center in our definition, and is therefore sensitive to spikes during merger events. Curves were smoothed using a moving average filter (270 Myr) to improve readability.
\label{fig:disk_radius_evolution}}
\end{figure}

\begin{figure}[ht!]
\includegraphics[width=0.48\textwidth]{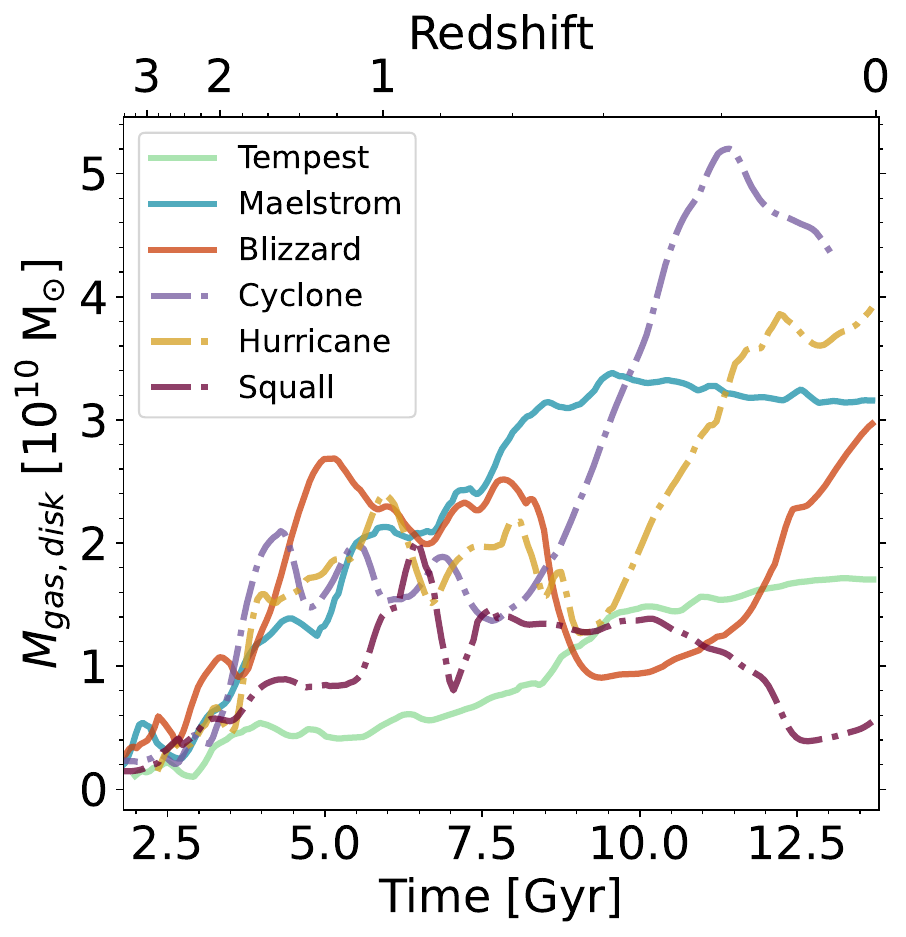}
\caption{
Evolution of the gas mass of each disk, based on our disk definition. Although overall baryon content increases more or less monotonically in all cases, some systems lose large amounts of gas, particularly at higher redshift. Overall trends seem uncorrelated with classification (denoted by line type). Curves were smoothed using a moving average filter (270 Myr) to improve readability.
\label{fig:disk_mass_evolution}}
\end{figure}

Fig.~\ref{fig:disk_radius_evolution} and Fig.~\ref{fig:disk_mass_evolution} show the evolution of how the maximum radial extent of each disk definition and the enclosed gas mass evolve over cosmological time. The physical size of each disk evolves similarly, regardless of whether it is \LessActive\ or \MoreActive\. The enclosed gas mass varies more significantly, but is not correlated with this classification.

 Tempest and Maelstrom, the two galaxies in the least populated environments, increase in size or less monotonically. Maelstrom grows continuously over time, while Tempest's size only changes significantly between $t\sim5$ Gyr ($z\sim1.3$) and $t\sim10$ Gyr ($z\sim0.3)$. Blizzard grows in a fashion similar to Maelstrom, except for a period around $t\sim9-10$ Gyr in which the disk shrinks, related to a merger/starburst event and the disruption of its polar ring. The change in gas mass roughly follows the change in size. Interestingly, despite continuing to increase in size at $z=0$, Maelstrom is steadily losing gas mass. This change is less than the increase in stellar mass during this period, \rev{which may indicate it is being locked up in star formation.}

The \MoreActive~galaxies show more disrupted size evolutions. Squall rapidly grows to a radius of $\sim$30 kpc at t$\sim6.5$ ($z\sim0.9$) Gyr, but shrinks after about 1 Gyr. While it recovers to this peak size, it never grows much larger, and its gas mass does not recover. This is due to a major merger causing an increase in disk size, followed by a starburst and a brief quenching event (see Appendix~\ref{sec:other_halo_time_evolutions}). Hurricane grows relatively slowly up to $t\sim$7.5 Gyr ($z\sim0.7$), after which it grows through a series of rapid size increases, followed by drops. Its mass evolution roughly follows this. Finally, Cyclone similarly starts growing more rapidly at $t\sim7.5$ Gyr ($z\sim0.7$), before shrinking and stabilizing at around $t=10$ Gyr ($z\sim0.3$). Its mass continues to increase, implying it is becoming increasingly compact.

\bibliography{hi_disks}{}
\bibliographystyle{aasjournalv7}



\end{document}